\documentclass[10pt]{amsart}
\usepackage{amsbsy,amssymb,amscd,amsfonts,latexsym,amstext,delarray,
amsmath,amsthm,graphicx}

\input xypic

\newtheorem{thm}{Theorem}[section]
\newtheorem{prop}[thm]{Proposition}
\newtheorem{cor}[thm]{Corollary}
\newtheorem{lem}[thm]{Lemma}

\newtheorem{defn}[thm]{Definition}
\newtheorem{rem}[thm]{Remark}

\numberwithin{equation}{section}

\def\bG{{\mathbb G}}

\def\bU{{\mathbb U}}

\def\A{{\mathbb A}}
\def\C{{\mathbb C}}

\def\N{{\mathbb N}}
\renewcommand{\P}{{\mathbb P}}
\def\Q{{\mathbb Q}}
\def\Z{{\mathbb Z}}
\def\R{{\mathbb R}}
\def\K{{\mathbb K}}
\def\m{{\mathfrak{m}}}
\def\cg{{\mathfrak{g}}}

\def\cC{{\mathcal C}}
\def\cD{{\mathcal D}}

\def\cF{{\mathcal F}}

\def\cH{{\mathcal H}}
\def\cI{{\mathcal I}}

\def\cK{{\mathcal K}}
\def\cL{{\mathcal L}}
\def\cM{{\mathcal M}}
\def\cN{{\mathcal N}}
\def\cO{{\mathcal O}}
\def\cP{{\mathcal P}}
\def\cQ{{\mathcal Q}}
\def\cR{{\mathcal R}}
\def\cS{{\mathcal S}}
\def\cT{{\mathcal T}}
\def\cU{{\mathcal U}}

\newcommand{\ie}{{\it i.e.\/}\ }
\newcommand{\eg}{{\it e.g.\/}\ }
\newcommand{\cf}{{\it cf.\/}\ }
\def\text{\hbox}

\def\codim{{\rm codim}}

\def\Ext{{\rm Ext}}

\def\Hom{{\rm Hom}}

\def\Lie{{\rm Lie}}
\def\ord{{\rm ord}}

\def\Res{{\rm Res}}

\def\SL{{\rm SL}}

\title{Motivic renormalization and singularities}
\author[Matilde Marcolli]{Matilde Marcolli}
\address{Department of Mathematics  \\
California Institute of Technology \\
Mail Code 253-37, 1200 E.California Blvd \\
Pasadena, CA 91125, USA} \email{matilde\@@caltech.edu}

\begin{document}
\maketitle

\begin{verse}{\em
Cos\`i tra questa \\
infinit\`a s'annega il pensier mio:\\
e 'l naufragar m'\`e dolce in questo mare.}\\
\smallskip
(Giacomo Leopardi, {\em L'infinito}, from the second handwritten version)
\end{verse}

\bigskip

{\em To Alain Connes, on his 60th birthday and many other occasions}

\bigskip

\begin{abstract} We consider parametric Feynman integrals and their
dimensional regularization from the point of view of differential
forms on hypersurface complements and the approach to mixed Hodge
structures via oscillatory integrals. We consider restrictions to
linear subspaces that slice the singular locus, to handle the 
presence of non-isolated singularities. In order to account for all possible choices 
of slicing, we encode this extra datum as an enrichment of the Hopf 
algebra of Feynman graphs. We introduce a new regularization
method for parametric Feynman integrals, which is based on Leray
coboundaries and, like dimensional regularization, replaces a
divergent integral with a Laurent series in a complex parameter. The
Connes--Kreimer formulation of renormalization can be applied to this
regularization method. We relate the dimensional regularization of
the Feynman integral to the Mellin transforms of certain
Gelfand--Leray forms and we show that, upon varying the external
momenta, the Feynman integrals for a given graph span a family of
subspaces in the cohomological Milnor fibration. We show how to pass
from regular singular Picard--Fuchs equations to irregular singular
flat equisingular connections. In the last section, which is more
speculative in nature, we propose a geometric model for
dimensional regularization in terms of logarithmic motives and 
motivic sheaves.
\end{abstract}

\tableofcontents

\section{Introduction}

We consider here perturbative quantum field theories governed by a 
Lagrangian, which in a Lorentzian metric of signature $(+1,-1,\ldots,-1)$
on the flat $D$-dimensional spacetime $\R^D$, is given in the
form
\begin{equation}\label{Lagr}
 \cL(\phi)= \frac 12 (\partial \phi)^2 - \frac{m^2}{2} \phi^2 -
\cL_{int}(\phi),
\end{equation}
where the interaction term $\cL_{int}(\phi)$ is polynomial in $\phi$
of degree at least three.
The corresponding action functional $S(\phi)=\int \cL(\phi) d^D x$
involves a single scalar field $\phi$. This is the simplest case,
considered in the work of Connes--Kreimer. 
Generalizations of the Connes--Kreimer
formalism for other theories have been developed more 
recently (see for instance
\cite{WvS} for the case of gauge theories), but for 
the purposes of the present 
paper we restrict our attention to scalar theories.

\smallskip

The purpose of this paper is to relate the approach to renormalization
of Connes--Kreimer \cite{CoKr}, via Birkhoff factorization of loops in the Lie
groups of characters of the Hopf algebra of Feynman graphs, and its
successive reformulation of Connes--Marcolli \cite{CoMa} in terms of
Galois theory of a category of flat equisingular connections with
irregular singularities, to the approach via parametric Feynman
integrals, periods of complements of graph hypersurfaces, and motives,
developed by Bloch--Esnault--Kreimer in \cite{BEK}, \cite{Blo}. 

\smallskip

The main approach we follow in this paper, in order to bridge 
between these two different approaches is a
formulation of the dimensionally regularized Feynman integrals in
terms of Mellin transforms of certain Gelfand--Leray forms, as in the
approach of Varchenko \cite{Var1}, \cite{Var2} to the theory of
singularities and asymptotic mixed Hodge structures on the
cohomological Milnor fibration, in terms of asymptotic properties of
oscillatory integrals. 

\smallskip

We deal with the fact that the graph hypersurfaces tend to
have non-isolated singularities by slicing the Feynman integral
along generic linear spaces of dimension at most equal to the 
codimension of 
the singular locus, using the same kind of techniques used in the
integral geometry of Radon transforms in projective spaces developed
by Gelfand--Gindikin--Graev \cite{GGG}. Since typically the singular
locus is rather large in dimension, the slices obtained in this way
will often be singular curves in $\P^2$ or singular surfaces in $\P^3$. 
Instead of
considering a single choice of a slicing, which would mean losing too
much information on the graph hypersurface, one considers all possible
choices and implements the data of the cutting linear space as part of
the Hopf algebra of graphs, much like what one does with the choice of
the external momenta, so that all possible choices are considered as
part of the structure. 

\smallskip

The formulation one obtains in this way, in terms of 
Gelfand--Leray forms, suggests a new method of regularization of
parametric Feynman integrals, which, as in the case of dimensional
regularization, replaces a divergent integral with a Laurent series in
a complex variable $\epsilon$, but which is defined using Leray
coboundaries to avoid the singular locus, by integrating around it
along the fibers of a circle bundle. We check that the formulation of
renormalization in terms of Hopf algebras and Birkhoff factorization
developed in Connes--Kreimer \cite{CoKr} applies without changes if 
one uses this new regularization method instead of the customary 
dimensional regularization. 

\smallskip

The interpretation of the dimensionally regularized Feynman integrals
as Mellin transforms of Gelfand--Leray forms provides a direct link
between Feynman integrals and the cohomological Milnor fibration. In
particular, we prove that, upon varying the external momenta and the
spacetime dimension $D\in \N$ in which the scalar theory is
considered, the corresponding Feynman integrals determine a family of
subspaces of the cohomological Milnor fibration, which inherit a Hodge
and a weight filtration from the asymptotic mixed Hodge structure of
Varchenko. It remains to be seen when this subspace recovers the full
Milnor fiber cohomology and/or when these induced filtrations still 
define a mixed Hodge structure. 

\smallskip

Another important question, in trying to compare the approaches of
\cite{CoMa} and \cite{BEK} is the use of irregular, as opposed to
regular singular, connections. In fact, from the point of view of
motives or mixed Hodge structures, what one expects to find is regular
singular connections. These appear naturally in the form of
Picard--Fuchs equations and Gauss--Manin connections. However, the
Galois theory approach to the classification of divergences in
perturbative quantum field theory developed in \cite{CoMa} relies
on the use of irregular singular connections and a 
form of the Riemann--Hilbert correspondence based on Ramis' wild
fundamental group. We reconcile these two approaches by showing that,
upon passing to Mellin transforms of solutions of a regular singular
Picard--Fuchs equation, one obtains solutions of differential equations 
with irregular singularities. More precisely, we first recall the
construction and properties of the irregular singular connections 
considered in \cite{CoMa} and the equisingularity condition that 
characterizes them. We then prove that solutions of the regular singular 
Picard--Fuchs equations at the singular points of a graph hypersurface
(sliced with a linear space of a suitable dimension so that
singularities are isolated) can be assembled to give rise to a
solution of a differential system of the type considered in
\cite{CoMa}, with irregular singularities and with coefficients in the
Lie algebra of the affine group scheme of the Hopf algebra of Feynman
graphs of the theory, suitably enriched to account for the choice of
the slicing of the Feynman integrals by linear spaces of the
appropriate dimension. 

\smallskip

Finally, we propose a motivic interpretation for dimensional
regularization, in terms of the logarithmic extensions of Tate motives
(the Kummer motives), and their pullbacks via the polynomial function
defining the graph hypersurface. This amounts to associating
to the Feynman graphs of a given scalar theory a subcategory of the
Arapura category of motivic sheaves of \cite{Arapura}. 
We expect that this may provide a way of interpreting
the relation between dimensionally regularized Feynman integrals and
cohomological Milnor fibrations in terms of a motivic version of the
Milnor fiber. We hope to relate, in this way, a motivic zeta function
associated to the resulting mixed motive with the dimensionally
regularized Feynman integral. 

\smallskip

An in depth study of parametric Feynman integrals in perturbative
renormalization and their relation to mixed Hodge structures was
carried out in very recent work of Bloch and Kreimer \cite{BloKr}.
 
\medskip

{\bf Acknowledgment.} Part of this work was carried out during a stay 
of the author at Florida State 
University, where the MAS6396 class provided a good sounding board for
various related topics. The author is partially supported by NSF 
grant DMS-0651925.

\section{Parametric Feynman integrals}\label{FeynSec}

In this section we recall the Feynman parametric formulation of the
momentum integrals associated to the Feynman graphs in the
perturbative expansion of a scalar field theory. We also recall the
Dimensional Regularization method and the form of the regularized
integrals. These are all well known techniques, but we review them
briefly for completeness. We also recall the explicit form of the 
graph polynomials $\Psi_\Gamma(t)$ and $P_\Gamma(t,p)$ and their 
properties, as well as the explicit mass scale dependence of the
dimensionally regularized Feynman integrals. 
Moreover, in \S \ref{IntProjSec} 
we give a reformulation of the Feynman integrals in terms of differential
forms on hypersurface complements in projective spaces. 

\subsection{Feynman parameters and algebraic varieties}\label{VarSec}

We recall briefly the method for the computation of Feynman 
integrals based on the
{\em parametric representation}. This is well known material 
in the physics literature,
see \eg \S 6-2-3 of \cite{ItZu}, \S 18 of \cite{BjDr}, 
and \S 6 of \cite{Naka}. 
However, since it is not part of the standard 
mathematician's toolbox, we prefer to 
spend a few words here recalling the basic ideas. 

The terms in the formal asymptotic expansion of functional integrals 
$$ \int \cO(\phi) e^{\frac{i}{\hbar} S(\phi)} \cD[\phi], $$
obtained by treating the interaction terms $S_{int}(\phi)=\int \cL_{int}(\phi) d^Dx$ 
as a perturbation, are labeled by Feynman graphs of the theory. 
The topology of 
these graphs is constrained by the requirement that the valence of each vertex
is equal to the degree of one of the monomials in the Lagrangian. The edges of
the graph are divided into internal lines, each connecting two vertices, and 
external lines, which are half-lines with one end attached to a vertex of the graph
and one open end. The order in the
expansion is given by the loop number of the graph, or by the number of
internal lines. Each external line carries 
a datum of an external momentum $p \in \R^D$ with a conservation law
\begin{equation}\label{extmom0}
 \sum_{e\in E_{ext}(\Gamma)} p_e =0, 
\end{equation}
where $E_{ext}(\Gamma)$ is the set of external edges of $\Gamma$.

We assume that all our graphs are one-particle-irreducible (1PI), \ie
that they cannot be disconnected by cutting a single internal edge. 

The Feynman rules assign to a Feynman graph a function
$U(\Gamma)=U(\Gamma,p_1\ldots,p_N)$ of the external momenta obtained
by integrating, over momentum variables $k_e$ assigned to each internal
edge of $\Gamma$, an expression involving propagators for each internal
line and momentum conservations at each vertex, in the form 
\begin{equation}\label{FeynmanU}
U(\Gamma)=\prod_{v\in V(\Gamma)} (2\pi)^D \lambda_v 
\int \frac{\delta(\sum_{i=1}^n \epsilon_{v,i} k_i +\sum_{j=1}^N \epsilon_{v,j} p_j)}
{q_1\cdots q_n} \,\, d^D k_1\cdots d^D k_n ,
\end{equation}
where $n=\# E_{int}(\Gamma)$, the number of internal edges in the graph, $N=\# E_{ext}(\Gamma)$, 
while $\lambda_v$ are the coupling constants (the coefficients of the term in the Lagrangian
of degree equal to the valence of the vertex $v\in V(\Gamma)$),
and $\epsilon_{v,e}$ is the incidence matrix
\begin{equation}\label{incidence}
\epsilon_{v,e}=\left\{ \begin{array}{rl} +1 & t(e)=v \\
-1 & s(e)=v \\
0 & \text{otherwise,} \end{array} \right.
\end{equation}
with $s(e)$ and $t(v)$ the source and target vertices of the oriented edge $e$.
In the following we drop the multiplicative constants $(2\pi)^D \lambda_v$ and
we concentrate on the remaining integral, which we still denote by $U(\Gamma)$.

The basic formula \eqref{FeynmanU} is dictated by the Feynman rules of the
given quantum field theory, which prescribe that the contribution of a Feynman graph 
to the perturbative expansion of the effective action is obtained as a product of 
an inverse propagator $q_e^{-1}$ for each edge $e$, in a corresponding momentum variable,
with linear relations among the momentum variables given by imposing
conservation laws at each vertex that ensure momentum conservation, and with each vertex 
contributing a multiplicative factor depending on coupling constants and powers of $(2\pi)$. 
These Feynman rules are modeled on similar series expansions of finite dimensional Gaussian
integrals.

The $q_i$, for $i=1,\ldots,n$ are the quadratic forms defining the free field propagator 
associated to the corresponding line in the graph, namely 
\begin{equation}\label{qs}
q_i(k)= k_i^2 - m^2 +i\epsilon \ \ \ \text{ or } \ \ \  q_i(k)=k_i^2 +m^2,
\end{equation}
respectively in the Lorentzian and in the Eucidean signature case.  In
the following, we work preferably in the Euclidean setting.   

We refer to $U(\Gamma)$ as the {\em unrenormalized Feynman integral}.
The parametric form of $U(\Gamma)$ is obtained by first introducing
the {\em Schwinger parameters}, using the identity 
$$ \frac{1}{q}=\int_0^\infty e^{-s q} ds. $$
This gives  the expression
\begin{equation}\label{Schwinger}
\frac{1}{q_1 \cdots q_n} =
\int_0^\infty \cdots \int_0^\infty e^{-(s_1 q_1 +\cdots + s_n q_n)}\,  \,\,ds_1 \cdots ds_n, 
\end{equation}
which is a special case of the more general identity
\begin{equation}\label{prodqkGamma}
\frac{1}{q_1^{k_1}\cdots q_n^{k_n}} = \frac{1}{\Gamma(k_1) \cdots \Gamma(k_n)} 
\int_0^\infty \cdots \int_0^\infty e^{-(s_1 q_1 +\cdots + s_n q_n)}\, s_1^{k_1-1} \cdots s_n ^{k_n-1}
 \,\,ds_1 \cdots ds_n.
\end{equation}
The Feynman parametric form is obtained from this expression by a
change of variables that replaces the Schwinger parameters $s_i \in
\R_+$ with new variable $t_i\in [0,1]$, by setting $s_i =S t_i$ with
$S=s_1+\cdots+s_n$. This gives 
\begin{equation}\label{prodqkSt} 
\frac{1}{q_1^{k_1}\cdots q_n^{k_n}} =
\frac{\Gamma(k_1+\cdots+k_n)}{\Gamma(k_1) \cdots \Gamma(k_n)}
\int_0^1 \cdots \int_0^1 \frac{t_1^{k_1-1} \cdots t_n^{k_n-1}\,
\delta(1-\sum_{i=1}^n t_i)}{(t_1 q_1 +\cdots+ t_n q_n)^{k_1+\cdots
+k_n}}  
 \,\,dt_1 \cdots dt_n,
\end{equation}
hence in particular one obtains
\begin{equation}\label{SchwingerFeynman}
\frac{1}{q_1 \cdots q_n} =(n-1)! \, \int \frac{\delta(1-\sum_{i=1}^n t_i)}{(t_1 q_1 + \cdots +
t_n q_n)^n} \,\, dt_1\cdots dt_n,
\end{equation}
as an integration in the Feynman parameters $t=(t_i)$ over the simplex 
\begin{equation}\label{simplex}
\Sigma= \{ t=(t_i)\in \R_+^n \,|\, \sum_i t_i =1 \}.
\end{equation}

Next one introduces a further change of variables involving another
matrix naturally associated to the graph, the circuit matrix
$\eta_{ik}$, defined in terms of an orientation of the edges $e_i \in E(\Gamma)$ and  
a choice of a basis for the first homology group, $l_k \in
H_1(\Gamma,\Z)$,  with $k=1,\ldots, \ell=b_1(\Gamma)$, by setting 
\begin{equation}\label{etaik}
\eta_{ik}=\left\{ \begin{array}{rl} +1 & \text{edge $e_i\in$ loop
$l_k$, same orientation} \\[2mm] -1 & \text{edge $e_i\in$ loop
$l_k$, reverse orientation} \\[2mm] 0 & \text{otherwise.} \end{array}\right.
\end{equation}
We also define $M_\Gamma(t)$ to be the matrix
\begin{equation}\label{MGamma}
(M_\Gamma)_{kr}(t)=\sum_{i=0}^n t_i \eta_{ik} \eta_{ir}.
\end{equation}
Notice that, while the matrix $M_\Gamma(t)$ depends on the choice of the
orientation of the edges and of the choice of a basis for the first homology
of $\Gamma$, the determinant $\det(M_\Gamma(t))$ is independent of both choices.

One then makes a change of variables in the quadratic forms $q_i$ of \eqref{qs}, by setting
\begin{equation}\label{varchangepi}
 k_i = u_i + \sum_{k=1}^\ell \eta_{ik} x_k, 
\end{equation} 
with the constraint 
\begin{equation}\label{uxeta} 
  \sum_{i=0}^n t_i u_i \eta_{ik} =0, 
\end{equation}
for all $k=1,\ldots,\ell$.  The momentum conservation relations 
$$ \sum_{i=1}^n \epsilon_{v,i} k_i +\sum_{j=1}^N \epsilon_{v,j} p_j =0 $$
of \eqref{FeynmanU} shows that the $u_i$ in \eqref{varchangepi} also satisfy
\begin{equation}\label{upepsilon}
\sum_{i=1}^n \epsilon_{v,i} u_i + \sum_{j=1}^N \epsilon_{v,j} p_j =0 .
\end{equation}
This uses the fact that the incidence matrix $\epsilon=(\epsilon_{v,e})$ and 
the circuit matrix $\eta=(\eta_{e,k})$ satisfy $\epsilon \eta=\sum_{e\in E(\Gamma)} 
\epsilon_{v,e} \eta_{e,k} =0$, \cf \cite{Naka}, \S 3.
The two equations \eqref{uxeta} and \eqref{upepsilon} are the analog
for momenta in Feynman graphs of the Kirchhoff laws of circuits,
respectively giving the conservation laws for the sum of voltage drops
along a loop in a circuit and of incoming currents at a vertex, with
momenta replacing currents and the Feynman parameters in the role of
resistances (\cite{BjDr}, \S 18). 

The $u_i$ are determined by \eqref{uxeta} and \eqref{upepsilon}, and one can write the term
$\sum_i t_i (u_i^2+m^2)$ in the form of a function of the Feynman
parameters $t$ and the external momenta $p$ of the form 
\begin{equation}\label{Vtp}
V_\Gamma (t,p):= p^\tau R_\Gamma (t) p + m^2, 
\end{equation} 
where we use the fact that $\sum_i t_i=1$. The $N\times N$-matrix $R(t)$, with $N=\# E_{ext}(\Gamma)$ 
is constructed out of another matrix associated to the graph. This is
obtained as follows (\cf \cite{ItZu}, \S 6-2-3). 
Let $D_\Gamma(t)$ denote the matrix
\begin{equation}\label{DGammat}
(D_\Gamma(t))_{v,v'}=\sum_{i=1}^n \epsilon_{v,i}\, \epsilon_{v',i}\, t_i^{-1},
\end{equation}
with $n=\# E_{int}(\Gamma)$ and with $\epsilon_{v,i}$ the incidence matrix as in \eqref{incidence}.
Then the quadratic form $p^\tau  R(t) p$ of \eqref{Vtp} has the form 
\begin{equation}\label{RGammat}
p^\tau R_\Gamma p = \sum_{v,v'} P_v (D_\Gamma(t)^{-1})_{v,v'} P_{v'}, 
\end{equation}
where 
\begin{equation}\label{Pv}
P_v=\sum_{e\in E_{ext}(\Gamma), t(e)=v} p_e
\end{equation}
is the sum of the incoming external momenta at the vertex $v$.

\medskip

Summarizing the previous discussion,
the result of the change of variables \eqref{varchangepi} is that we can rewrite the original
Feynman integral \eqref{FeynmanU} in the following form.

\begin{lem}\label{paramform}
For $n-D\ell/2 >0$, the Feynman integral \eqref{FeynmanU} can be
written, after the change of variables \eqref{varchangepi}, in the form
\begin{equation}\label{UGammaParam}
 \int_{\Sigma} \frac{\delta(1-\sum_i t_i)}{\det(M_\Gamma(t))^{D/2}
V_\Gamma(t,p)^{n-D\ell/2}}\,\, dt_1\cdots dt_n, 
\end{equation}
up to a multiplicative constant. 
\end{lem}

\proof This follows \cite{BjDr}, \S 18 and \cite{ItZu} p.376. 
First recall the well known identity for the Gaussian integral
\begin{equation}\label{GaussianA}
\int e^{-\frac{1}{2} x^\tau A x} d^D x_1 \cdots d^D x_\ell = \frac{(2\pi)^{D\ell/2}}{\det(A)^{D/2}},
\end{equation}
for $A$ an $\ell\times \ell$ real symmetric matrix. We then have
$$ \frac{1}{(4\pi)^{D\ell/2}} \int e^{- x^\tau A x} d^D x_1 \cdots d^D
x_\ell = \det(A)^{-D/2}. $$
With the change of variable \eqref{varchangepi} and the conditions
\eqref{uxeta} and \eqref{upepsilon}, one can rewrite the integral
$U(\Gamma)$ of \eqref{FeynmanU} in the form
\begin{equation}\label{preDRint}
U(\Gamma)=  \int_{\R_+^n} 
e^{-V_\Gamma (t,p)} \left( \int    e^{-x^\tau M_\Gamma(t) x}
d^Dx_i \cdots d^D x_\ell \right) \,\, dt_1 \cdots dt_n ,   
\end{equation}
with $\ell=b_1(\Gamma)$ is the number of loops in the graph.
After performing the Gaussian integration and rewriting the expression
in the external momenta as described above in the form \eqref{Vtp} and
\eqref{RGammat}, this becomes of the form
\begin{equation}\label{expint}
U(\Gamma)= (4\pi)^{-\ell D/2} \int_{\R_+^n} 
\frac{e^{-V_\Gamma(t,p)}}{\Psi_\Gamma(t)^{D/2}} \,\, dt_1 \cdots dt_n,    
\end{equation}
with 
\begin{equation}\label{PsiGammaDet}
\Psi_\Gamma(t)=\det M_\Gamma(t).
\end{equation}
Then using the identity $1=\int_0^\infty d\lambda\, \delta(\lambda -\sum_{i=1}^n t_i)$ and scaling
$t_i \mapsto t_i \lambda$, one rewrites \eqref{expint} in the form
\begin{equation}\label{expint2}
U(\Gamma)= (4\pi)^{-\ell D/2} \int_0^\infty  \left(  \int_{[0,1]^n} \delta(1-\sum_i t_i) 
\frac{e^{-\lambda V_\Gamma(t,p)}}{\Psi_\Gamma(t)^{D/2}} \, dt_1 \cdots dt_n \right) 
\lambda^{n-D\ell/2}\frac{d\lambda}{\lambda}.
\end{equation}
Using again the special form
\begin{equation}\label{Vdlambda}
V_\Gamma^{-n+ D\ell/2} = \frac{1}{\Gamma(n-D\ell/2)} \int_0^\infty e^{-\lambda V_\Gamma} 
\lambda^{n-D\ell/2-1}\, d\lambda 
\end{equation}
of the general identity \eqref{prodqkGamma}, one then obtains the parametric form
\begin{equation}\label{expint3}
U(\Gamma)=   \frac{\Gamma(n-D\ell/2)}{(4\pi)^{\ell D/2}}  \int_{[0,1]^n} \frac{\delta(1-\sum_i t_i)}
{\Psi_\Gamma(t)^{D/2}V_\Gamma(t,p)^{n- D\ell/2}} \, dt_1 \cdots dt_n. 
\end{equation}
The condition $n-D\ell/2 = \ell (1-D/2) +\# V(\Gamma)-1 >0$ ensures
the convergence at $\lambda=0$ of the integral \eqref{Vdlambda}.
\endproof

\medskip

A graph is said to be {\em log divergent} if $n=D\ell/2$, in which case the Feynman integral 
reduces to the simpler form 
\begin{equation}\label{logdivGamma}
 \int_{\Sigma} \frac{\omega}{\det(M_\Gamma(t))^{D/2}}  ,
\end{equation}
with $\omega =\delta(1-\sum_i t_i) dt_1 \cdots dt_n$ the volume form on the 
simplex $\Sigma$ defined by the integration \eqref{expint3}.

\medskip

\begin{rem}\label{Qpi}{\rm
For the purpose of establishing relations between values of Feynman integrals and periods of motives, 
it is important to check that the multiplicative constant one is neglecting in passing from 
\eqref{FeynmanU} to the parametric form \eqref{UGammaParam} in fact belongs to $\Q(\pi)$, 
\cf \cite{BEK}.
In \eqref{expint3} one sees in fact that the multiplicative constant is of the form
$\Gamma(n-D\ell/2)(4\pi)^{-\ell D/2}$. This either gives a divergent factor, at the
poles of the Gamma function, in which case one considers the residue, or else, when
convergent, it gives a multiplicative factor in $\Q(\pi)$.}
\end{rem}

\medskip

The function $\Psi_\Gamma(t)=\det(M_\Gamma(t))$ has an equivalent
expression in terms of the  
connectivity of the graph $\Gamma$ as the polynomial (see \cite{ItZu},
\S 6-2-3 and \cite{Naka} \S 1.3-2) 
\begin{equation}\label{PhiGamma}
\Psi_\Gamma(t) = \sum_{S} \prod_{e\in S} t_e,
\end{equation}
where $S$ ranges over all the sets $S \subset E_{int}(\Gamma)$ of
$\ell=b_1(\Gamma)$ internal edges of $\Gamma$,  
such that the removal of all the edges in $S$ leaves a connected
graph. This can be equivalently formulated  
in terms of spanning trees of the graph $\Gamma$ (\cite{Naka} \S 1.3),
\ie $\Psi_\Gamma(t)$  
is given by the Kirchhoff polynomial
\begin{equation}\label{PsiGamma}
\Psi_\Gamma(t)= \sum_{T} \prod_{e \notin T} t_e,
\end{equation}
with the sum over spanning trees $T$ of the graph. Each spanning tree,
in fact, has $\# V-1$ edges and is the 
complement of a cut-set $S$. 

\medskip

\begin{lem}\label{homogPsi}
The graph polynomial $\Psi_\Gamma$ is a homogeneous polynomial of degree 
\begin{equation}\label{degPsiGamma}
\deg \Psi_\Gamma = b_1(\Gamma).
\end{equation}
In the {\em massless case} with $m=0$, the function $V_\Gamma(t,p)$, for fixed $p$, 
is homogeneous of degree one and given by the ratio of a homogeneous polynomial
$P_\Gamma(t,p)$ by $\Psi_\Gamma(t)$.
\end{lem}

\proof
We have $\deg \Psi_\Gamma = \# E(\Gamma) - \# E(T)$, where $\# E(T)=\# V(\Gamma) -1$ is
the nuber of edges in a (hence any) spanning tree, hence from the Euler characteristic formula 
$\# V(\Gamma)-\# E(\Gamma) = 1- b_1(\Gamma)$ we get \eqref{degPsiGamma}. 
We write the polynomial $V_\Gamma(t,p)=p^\tau R_\Gamma(t) p + \sum_i
t_i m^2$. In the massless case, using the reformulation given in (6-87) and (6-88) of
\cite{ItZu}, p.297, we rewrite the function $V_\Gamma(t,p)$ in the form of the ratio 
\begin{equation}\label{ratioVGamma}
V_\Gamma(t,p) = \frac{P_\Gamma(t,p)}{\Psi_\Gamma(t)}
\end{equation}
of a homogeneous polynomial $P_\Gamma$ of degree $\ell+1=b_1(\Gamma)+1$, divided by the
polynomial $\Psi_\Gamma$, which is homogeneous of degree
$b_1(\Gamma)$. In fact, we have (\cite{ItZu}, \S 6-2-3)
\begin{equation}\label{PGammapt}
P_\Gamma(p,t) = \sum_{C\subset \Gamma} s_C \prod_{e\in C} t_e, 
\end{equation}
where the sum is over the cut-sets $C\subset \Gamma$, \ie the
collections of $b_1(\Gamma)+1$ edges that divide the graph $\Gamma$ in
exactly two connected components $\Gamma_1\cup \Gamma_2$. 
The coefficient $s_C$ is a function of the external momenta attached 
to the vertices in either one of the two components
\begin{equation}\label{sCcoeff}
s_C = \left(\sum_{v\in V(\Gamma_1)} P_v\right)^2 = \left(\sum_{v\in
V(\Gamma_2)} P_v\right)^2, 
\end{equation}
where the $P_v$ are defined as in \eqref{Pv}, as the sum of the
incoming external momenta (see \cite{ItZu}, (6-87) and (6-88)).
\endproof

\smallskip

In the following, we work under the following assumption on the graph $\Gamma$.

\begin{defn}\label{genericp}
A 1PI graph $\Gamma$ satisfies the {\em generic condition} on the
external momenta if, for $p$ in a dense open set in the space of
external momenta, the polynomials $P_\Gamma(t,p)$ and $\Psi_\Gamma(t)$
have no common factor. 
\end{defn}

To understand better the nature of this condition, it is useful to 
reformulate the polynomial $P_\Gamma(t,p)$ of \eqref{PGammapt} in terms of 
spanning trees of the graph. One has, in the case where $m=0$, 
\begin{equation}\label{treesPGamma}
P_\Gamma(p,t) =\sum_T \sum_{e'\in T} s_{T,e'} \, t_{e'} \, \prod_{e\in
T^c} t_e,
\end{equation} 
where $s_{T,e'}=s_C$ for the cut-set $C=T^c \cup \{ e' \}$.

The parameterizing space of the external momenta is the hyperplane in
the affine space $\A^{D\cdot\#E_{ext}(\Gamma)}$ obtained by imposing the 
conservation law
\begin{equation}\label{conslawpe}
\sum_{e\in E_{ext}(\Gamma)} p_e =0.
\end{equation}

Thus, the simplest possible configuration of external momenta is the
one where one puts all the external momenta to zero, except for a pair 
$p_{e_1}=p=-p_{e_2}$ associated to a choice of a pair of external
edges $\{ e_1, e_2\}\subset E_{ext}(\Gamma)$. Let $v_i$ be the unique
vertex attached to the external edge $e_i$ of the chosen pair. 
We then have, in this case, $P_{v_1}=p=-P_{v_2}$. Upon writing the
polynomial $P_\Gamma(t,p)$ in the form \eqref{treesPGamma}, we obtain
in this case 
\begin{equation}\label{vwPGamma}
P_\Gamma(p,t) =p^2 \sum_T (\sum_{e'\in T_{v_1,v_2}} t_{e'}) \prod_{e
\notin T} t_e,
\end{equation}
where $T_{v_1,v_2} \subset T$ is the unique path in $T$ without
backtrackings connecting the vertices $v_1$ and $v_2$. 
We use \eqref{sCcoeff} to get $s_C=p^2$ for all the nonzero terms in
this \eqref{vwPGamma}. These are all the terms that corresponds to cut
sets $C$ such that the vertices $v_1$ and $v_2$ belong to different
components. These cut-sets consist of the complement of a spanning
tree $T$ and an edge of $T_{v_1,v_2}$.

In the following we will make use of the notation 
\begin{equation}\label{hTpolyn}
L_T(t)=p^2 \sum_{e\in T_{v_1,v_2}} t_e
\end{equation}
for the linear functions in \eqref{vwPGamma}.

If the polynomial $\Psi_\Gamma(t)$ of \eqref{PsiGamma} divides
\eqref{vwPGamma}, one has 
$$ P_\Gamma(p,t) = \Psi_\Gamma(t)\cdot L(t), $$
for a degree one polynomial $L(t)$, which gives
$$ \sum_T (L_T(t)-L(t)) \prod_{e \notin T} t_e \equiv 0, $$
for all $t$. One then sees, for example, that the 1PI condition on the
graph $\Gamma$ is necessary in order to have the condition of Definition
\ref{genericp}. In fact, for a graph that is not 1PI, one may be able
to find vertices and momenta as above such that the degree one
polynomials $L_T(t)$ are all equal to the same $L(t)$. Generally,
the validity of the condition of Definition \ref{genericp} can be
checked algorithmically for a given graph.

\smallskip

One does not need to assume the condition of Definition
\ref{genericp}. However, several of our formulae become more
complicated if we allow the case where the polynomials $\Psi_\Gamma$
and $P_\Gamma(t,p)$ have common factors. Thus, for our purposes we
assume to work under the hypothesis that the ``generic condition on
the external momenta'' holds.

\smallskip

\begin{defn} \label{graphXGamma}
The affine graph hypersurface $\hat X_\Gamma$ is the zero locus of the Kirchhoff polynomial
\begin{equation}\label{XGamma}
\hat X_\Gamma =\{ t\in \A^n \,: \, \Psi_\Gamma(t)=0 \},
\end{equation}
with $n=\# E_{int}(\Gamma)$.
The locus of zeros of the polynomial $P_\Gamma(t,p)$, for fixed external momenta $p$,
also defines a hypersurface
\begin{equation}\label{YGamma}
\hat Y_\Gamma= \hat Y_\Gamma(p) :=\{ t \in \A^n \,|\, P_\Gamma(t,p)=0 \}.
\end{equation}
Since both $\Psi_\Gamma(t)$ and $P_\Gamma(t,p)$ are homogeneous polynomials in $t$,
we can consider corresponding projective hypersurfaces
\begin{equation}\label{XGammaProj}
X_\Gamma =\{ t=(t_1:\cdots:t_n)\in \P^{n-1} \,: \, \Psi_\Gamma(t)=0 \}
\end{equation}
of degree $b_1(\Gamma)$ and 
\begin{equation}\label{YGammaProj}
Y_\Gamma= Y_\Gamma(p) :=\{ t =(t_1:\cdots:t_n)\in \P^{n-1} \,|\, P_\Gamma(t,p)=0 \}.
\end{equation}
of degree $b_1(\Gamma)+1$.
\end{defn}

\medskip

In the case of log divergent graphs, or of arbitrary graphs in the
range with sufficiently large spacetime dimension $D$ (\ie for $D$ satisfying 
$-n + D\ell/2 \geq 0$, with $n=\# E_{int}(\Gamma)$ and $\ell=b_1(\Gamma)$),
the possible divergences of the
Feynman integral $U(\Gamma)$ depend on the intersection of the domain
of integration $\Sigma$ with the graph hypersurface $\hat X_\Gamma$ in
$\P^{n-1}$. Notice that the intersections $\Sigma \cap \hat X_\Gamma$
can only happen on the boundary $\partial \Sigma$, as in the interior
of $\Sigma$ the polynomial $\Psi_\Gamma$ takes strictly positive real
values. 
See \cite{BEK} and \cite{Blo} for a detailed analysis of
this case and for its motivic interpretation. More generally, for
non-log-divergent integrals of the form \eqref{UGammaParam}, outside 
of the range $-n + D\ell/2 \geq 0$, the singularities of the integral
also involve the intersections of the hypersurfaces $\hat Y_\Gamma(t,p)$
with the domain of integration $\Sigma$. This case requires in general
a more detailed analysis, as in this case some of the intersections may 
also appear away from the boundary of $\Sigma$, depending on the values
of the external momenta $p$, see \eg \cite{BjDr}, \S 18.

\subsection{Dimensional Regularization}\label{DimRegSec}

One of the main problems that emerged in the historic development of
perturbative quantum field theory is how to ``cure" the divergences
that occur systematically in the Feynman integrals \eqref{FeynmanU},
\ie the problem of renormalization.  
Usually this is treated by choosing a {\em regularization} method,
combined with a {\em renormalization} procedure. Regularization replaces a
divergent integral  \eqref{FeynmanU} with a function of additional
parameters that happens to have a pole or singularity at the special
value of the parameter that corresponds to the original integral, but which
is otherwise well defined and finite at nearby values of the parameter. 
Renormalization, on the other hand, gives a method for extracting
finite values from the regularized expressions in a way that is
consistent with the combinatorics of nested subdivergences, \ie
subgraphs of graphs with divergent Feynman integrals, which themselves
contribute divergences. 

\smallskip

The Connes--Kreimer theory \cite{CoKr} uses the regularization
method known as {\em dimensional regularization and minimal
subtraction}, combined with the renormalization procedure of
Bogoliubov--Parasiuk--Hepp--Zimmermann (BPHZ). It was later shown (see
\eg \cite{EG}) that the main results of Connes--Kreimer may be
applied to other regularization procedures, as long as the ``subtraction of
infinities" can be formulated in terms of a Rota--Baxter operator. The
projection of a Laurent series onto its polar part is an example of
such an operator, which corresponds to the ``minimal subtraction"
case. Using this more general formulation, it possible to extend the
Connes--Kreimer theory to other regularization methods, which makes
it possible, for instance, to extend it to the case of curved backgrounds
as in \cite{Agarwala}.
We concentrate here on the Dimensional Regularization and Minimal
Subtraction procedure. In fact, our purpose is to compare the approach 
to motives and renormalization of \cite{CoMa} with the one of \cite{BEK}, 
and we prefer to remain close to the formulation given in \cite{CoMa} 
using DimReg. 

\smallskip

Dimensional Regularization consists of formally extending the usual
Gaussian integration \eqref{GaussianA} from the case of integer
dimension $D\in \N$ to the case of a ``complexified dimension" $z\in
\C$, in a small neighborhood of $z=0$, by setting
\begin{equation}\label{DimRegdet}
\int e^{-\frac{1}{2} x^\tau A x} d^{D+z} x_1 \cdots d^{D+z} x_\ell :=
\frac{(2\pi)^{(D+z)\ell/2}}{\det(A)^{(D+z)/2}}, 
\end{equation}
This results is the analytic continuation of the parametric Feynman
integral formulae \eqref{expint}, \eqref{expint2}, \eqref{expint3} to
complex values of the dimension $D$. 

\smallskip

\begin{lem}\label{DRlem}
Upon replacing the integer dimension $D$ by a complexified
dimension $D\mapsto D+z$, with $z\in \Delta^*$ a small punctured disk
around $z=0$, the integral \eqref{preDRint} becomes of the form 
\begin{equation}\label{DRint}
U(\Gamma)(z)=\frac{\Gamma(n-\frac{(D+z)\ell}{2})}{(4\pi)^{\frac{\ell (D+z)}{2}}}  
\int_{[0,1]^n} \frac{\delta(1-\sum_i t_i)}
{\Psi_\Gamma(t)^{(D+z)/2}V_\Gamma(t,p)^{n- (D+z)\ell/2}} \, dt_1 \cdots dt_n. 
\end{equation}
\end{lem}

\proof One uses the same argument of Lemma \ref{paramform}, but using 
\eqref{DimRegdet} instead of \eqref{GaussianA} in
\eqref{preDRint}. This gives
\begin{equation}\label{expintDR}
U(\Gamma)= (4\pi)^{-\ell (D+z)/2} \int_{\R_+^n} 
\frac{e^{-V_\Gamma(t,p)}}{\Psi_\Gamma(t)^{(D+z)/2}} \,\, dt_1 \cdots dt_n,    
\end{equation}
We then use the same argument as in Lemma \ref{paramform} to write
this in the form
\begin{equation}\label{expint2DR}
U(\Gamma)= (4\pi)^{-\ell (D+z)/2} \int_0^\infty  \left(  \int_{[0,1]^n} \delta(1-\sum_i t_i) 
\frac{e^{-\lambda V_\Gamma(t,p)}}{\Psi_\Gamma(t)^{(D+z)/2}} \, dt_1 \cdots dt_n \right) 
\lambda^{n-(D+z)\ell/2}\frac{d\lambda}{\lambda}
\end{equation}
and we use
\begin{equation}\label{VdlambdaDR}
V_\Gamma^{-n+ (D+z)\ell/2} = \frac{1}{\Gamma(n-(D+z)\ell/2)} \int_0^\infty e^{-\lambda V_\Gamma} 
\lambda^{n-(D+z)\ell/2-1}\, d\lambda 
\end{equation}
to obtain \eqref{DRint}. One recovers the parametric form
\eqref{UGammaParam} from \eqref{DimRegdet}.  
\endproof

\medskip

\subsection{Mass scale dependence}\label{MassSec}

It is well known that, when one regularizes the integrals $U(\Gamma)$
using dimensional regularization, as recalled above, one introduces an
explicit dependence on the mass scale, which plays a very important
role in the renormalization process and is the source of the
nontrivial action of the renormalization group (see \cite{Collins}, \cite{CoKr},
\cite{CoMa}, \cite{CoMa-book}).

The source of the mass scale dependence is the fact that, in order to
maintain the physical units, the integral \eqref{DimRegdet} should in
fact be written in the form
\begin{equation}\label{muDimRegdet}
\int e^{-\frac{1}{2} x^\tau A x} \mu^{-z} d^{D+z} x_1 \cdots \mu^{-z}
d^{D+z} x_\ell := \mu^{-z\ell} \frac{(2\pi)^{(D+z)\ell/2}}{\det(A)^{(D+z)/2}},
\end{equation}
where $\mu$ has the physical units of a mass (energy), so that the
$\mu^{-z} d^{D+z} x_i$ still have the same physical units as the
original $d^D x_i$ (see \cite{Collins}).

\begin{lem}\label{massUGammaz}
The dimensional regularization $U(\Gamma)(z)$ of \eqref{DRint} depends
on the mass scale $\mu$ in the form
\begin{equation}\label{muDRint}
U_\mu(\Gamma)(z)=\mu^{-z\ell}\frac{\Gamma(n-\frac{(D+z)\ell}{2})}
{(4\pi)^{\frac{\ell (D+z)}{2}}}  \int_{[0,1]^n} \frac{\delta(1-\sum_i
t_i) dt_1 \cdots dt_n}
{\Psi_\Gamma(t)^{\frac{(D+z)}{2}}V_\Gamma(t,p)^{n- \frac{(D+z)\ell}{2}}} \, . 
\end{equation}
\end{lem}

\proof In the derivation of the parametric form of the 
Feynman integral with dimensional regularization, we see that we
have in \eqref{expintDR} a mass scale dependence
\begin{equation}\label{expintDRmu}
U_\mu(\Gamma)(z)= (4\pi)^{-\ell (D+z)/2} \mu^{-z\ell} \int_{\R_+^n} 
\frac{e^{-V_\Gamma(t,p)}}{\Psi_\Gamma(t)^{(D+z)/2}} \,\, dt_1 \cdots dt_n.    
\end{equation}
The rest of the argument of Lemma \ref{DRlem} is unchanged. In
particular, no further $\mu$ dependence is introduced by the 
term in $V_\Gamma(t,p)$, so that we obtain \eqref{muDRint}.
\endproof

\medskip

\subsection{Integrals on projective spaces}\label{IntProjSec}

As remarked above, due to the homogeneity of the polynomials
$\Psi_\Gamma$ and $P_\Gamma$, it is natural to regard the graph hypersurfaces
as projective hypersurfaces $X_\Gamma$ and $Y_\Gamma$ in $\P^{n-1}$, with $n$ the
number of internal lines of the graph $\Gamma$. Thus, we want to think
of the parametric Feynman integrals as being computed in
projective space. 

\smallskip

In order to reformulate in projective space $\P^{n-1}$ integrals originally
defined in affine space $\A^n$, one needs to work with the projective analog
(\cf \cite{GGG}, \S II) of the volume form 
$$ \omega_n=dt_1 \wedge \cdots \wedge dt_n . $$  This is given by the form
\begin{equation}\label{volP}
\Omega = \sum_{i=1}^n (-1)^{i+1} t_i \, dt_1 \wedge \cdots \wedge \widehat{dt_i}
\wedge \cdots \wedge dt_n . 
\end{equation}
The relation between the volume form $dt_1 \wedge \cdots \wedge dt_n$
and the homogeneous form $\Omega$ of degree $n$ of \eqref{volP} is
given by (\cf \cite{Dimca}, p.180)
\begin{equation}\label{volDeltaOmega}
\Omega=\Delta(\omega_n), 
\end{equation}
where $\Delta: \Omega^k \to \Omega^{k-1}$ is the operator of
contraction with the Euler vector field 
\begin{equation}\label{Eulervec}
E=\sum_i t_i
\frac{\partial}{\partial t_i},
\end{equation}
\begin{equation}\label{EulerDelta}
 \Delta(\omega)(v_1,\cdots,v_{k-1})=
\omega(E,v_1,\cdots,v_{k-1}). 
\end{equation}

\smallskip

In the parametric Feynman integrals, we consider as region of
definition of the integrand (in the log divergent case, or in the case
of integrals in the range $-n +D\ell/2\geq 0$) the hypersurface complement
\begin{equation}\label{hypcomplA}
\cD(\Psi_\Gamma) =\{ t\in \A^n \,|\, \Psi_\Gamma(t)\neq 0 \} =\A^n
\smallsetminus \hat X_\Gamma,
\end{equation}
while in the formulation \eqref{expint3} outside of the range $-n+D\ell/2\geq 0$, 
we also need to avoid the second hypersurface $\hat Y_\Gamma$ defined by the 
vanishing of $P_\Gamma$ (for assigned external momenta), as in \eqref{YGamma}.
In this case the domain of definition of the integrand is
\begin{equation}\label{hypcomplAV}
\begin{array}{rl}
\cD(\Psi_\Gamma,P_\Gamma) = & \{ t\in \A^n \,|\, \Psi_\Gamma(t)\neq 0
\text{ and } P_\Gamma(t,p) \neq 0 \} \\[2mm]  = & \cD(\Psi_\Gamma)\cap
\cD(P_\Gamma) = \A^n \smallsetminus (\hat X_\Gamma \cup \hat Y_\Gamma).
\end{array}
\end{equation}

\smallskip

Let $\cU(\Psi_\Gamma)$ and $\cU(\Psi_\Gamma, P_\Gamma)$ denote the
corresponding hypersurface complements in projective space, namely
\begin{equation}\label{hypcomplP}
\begin{array}{rl}
\cU(\Psi_\Gamma) = & \{ t\in \P^{n-1} \,|\, \Psi_\Gamma(t)\neq 0 \} =\P^{n-1}
\smallsetminus X_\Gamma \\[3mm]
\cU(\Psi_\Gamma,P_\Gamma) = & \{ t\in \P^{n-1} \,|\, \Psi_\Gamma(t)\neq 0
\text{ and } P_\Gamma(t,p) \neq 0 \} \\[2mm]  = & \cU(\Psi_\Gamma)\cap
\cU(P_\Gamma) = \P^{n-1} \smallsetminus (X_\Gamma \cup Y_\Gamma).
\end{array}
\end{equation}

\smallskip

As we see in more detail in \eqref{fnDell} and Proposition
\ref{FeynDelta} below, in both the affine and the projective case, we can describe
$\cD(\Psi_\Gamma,P_\Gamma)$ and $\cU(\Psi_\Gamma,P_\Gamma)$ as
hypersurface complements, by identifying $X_\Gamma \cup Y_\Gamma$ with
the hypersurface defined by the vanishing of a homogeneous polynomial
given by a product $\Psi_\Gamma^{n_1}
\cdot P_\Gamma^{n_2}$, a homogeneous polynomial of degree $n_1
b_1(\Gamma)+ n_2(b_1(\Gamma)+1)$, where the component hypersurfaces $X_\Gamma$
and $Y_\Gamma$ are counted with multiplicities $n_1$ and $n_2$. These
multiplicities depend on the number of edges and loops of the graph
and on the spacetime dimension, and are defined more precisely in
\eqref{fnDell} below.
Thus, in the following, wherever needed, we write $\cD(\Psi_\Gamma,P_\Gamma)=
\cD(f)$ and $\cU(\Psi_\Gamma,P_\Gamma)=\cU(f)$, with $f=\Psi_\Gamma^{n_1}
\cdot P_\Gamma^{n_2}$, as in the various cases of \eqref{fnDell} below.

\smallskip

We introduce here some notation that will be useful in the following (\cf
\cite{Dimca}, p.177).
Let $\cR=\C[t_1,\ldots,t_n]$ be the ring of polynomials of $\A^n$. Let
$\cR_m$ denotes the subset of homogeneous polynomials of degree
$m$. Similarly, let $\Omega^k$ denote the $\cR$-module of $k$-forms on
$\A^n$ and let $\Omega^k_m$ denote the subset of $k$-forms that are
homogeneous of degree $m$. 

\smallskip

We recall the following general fact (\cf \cite{Dimca}, p.178) about
hypersurface complements. Let $\pi : \A^n\smallsetminus\{0\} \to \P^{n-1}$ be the
standard projection $t=(t_1,\ldots,t_n)\mapsto t=(t_1:\cdots :t_n)$. 
Suppose given a homogeneous polynomial function
$f$ on $\A^n$ of degree $\deg(f)$. Let $\cD(f)\subset \A^n$ and
$\cU(f)\subset \P^{n-1}$ be the hypersurface complements, \ie the
complements, in $\A^n$ and $\P^{n-1}$ respectively, of the locus of
zeros $X_f=\{ t\,|\, f(t)=0 \}$. 
With the notation introduced here above, we can always write a form 
$\omega \in \Omega^k(\cD(f))$ as
\begin{equation}\label{formfm}
\omega = \frac{\eta}{f^m}, \ \ \ \text{ with } \ \ \eta\in \Omega^k_{m\deg(f)}.
\end{equation}

We then have the following characterization of the pullback along
$\pi: \cD(f)\to \cU(f)$ of forms on $\cU(f)$ (see \cite{Dimca}, p.180
and \cite{Dolg}). Given $\omega \in \Omega^k(\cU(f))$, the pullback
$\pi^*(\omega) \in \Omega^k(\cD(f))$ is characterized by the
properties of being invariant under the $\bG_m$
action on $\A^n\smallsetminus \{0\}$ and of satisfying $\Delta(\pi^*(\omega))=0$, where
$\Delta$ is the contraction \eqref{EulerDelta} with the Euler vector field
$E$ of \eqref{Eulervec}. Thus, since the sequence 
$$ 0\to\Omega^n\stackrel{\Delta}{\to} \Omega^{n-1}
\stackrel{\Delta}{\to} \cdots \Omega^1 \stackrel{\Delta}{\to} \Omega^0
\to 0 $$
is exact at all but the last term, one can write
\begin{equation}\label{pistarforms}
 \pi^*(\omega) =\frac{\Delta(\eta)}{f^m}, \ \ \text{ with } \ \ 
\eta \in \Omega^k_{m\deg(f)}. 
\end{equation}
Thus, in particular, any $(n-1)$-form on $\cU(f)\subset \P^{n-1}$
can be written as
\begin{equation}\label{formUf}
\frac{h \Omega}{f^m}, \ \ \text{ with } \ \ h\in \cR_{m\deg(f)-n}
\end{equation}
and with $\Omega =\Delta(dt_1\wedge\cdots\wedge dt_n)$ the $(n-1)$-form
\eqref{volP}, homogeneous of degree $n$. 

\medskip

\begin{prop}\label{intSigmaprop}
Let $\omega \in \Omega^k_{m \deg(f)}$ be a closed $k$-form, which is
homogeneous of degree $m \deg(f)$, and consider the form $\omega /f^m$
on $\A^n$. Let $\Sigma \subset \A^n\smallsetminus \{0\}$ be a
$k$-dimensional domain with boundary $\partial \Sigma \neq \emptyset$.
Then the integration of $\omega /f^m$ on $\Sigma$ satisfies
\begin{equation}\label{intSigma}
m \deg(f) \, \int_\Sigma \frac{\omega}{f^m} = \int_{\partial \Sigma}
\frac{\Delta(\omega)}{f^m}  + 
\int_{\Sigma} df \wedge \frac{\Delta(\omega)}{f^{m+1}}. 
\end{equation}
\end{prop}

\proof Recall that we have (\cite{Dimca}, \cite{Dolg})
\begin{equation}\label{dDeltam}
d \left(\frac{\Delta(\omega)}{f^m}\right) = - \frac{\Delta(d_f \omega)}{f^{m+1}},
\end{equation}
where, for a form $\omega$ that is homogeneous of degree $m \deg(f)$,
\begin{equation}\label{dfomega}
d_f \omega = f\, d\omega - m\, df \wedge \omega. 
\end{equation}

Thus, we have
\begin{equation}\label{dDeltam2}
 d \left(\frac{\Delta(\omega)}{f^m}\right) = - \frac{\Delta(d\omega)}{f^m} + m
\frac{\Delta(df \wedge \omega)}{f^{m+1}}. 
\end{equation}
Since the form $\omega$ is closed, $d\omega=0$, and we have 
\begin{equation}\label{Deltadf}
\Delta(df \wedge \omega) = \deg(f) \, f\, \omega - df \wedge
\Delta(\omega), 
\end{equation}
we obtain from the above
\begin{equation}\label{dDeltam3}
d \left(\frac{\Delta(\omega)}{f^m}\right) = m \deg(f) \frac{\omega}{f^m} - \frac{df
\wedge \Delta(\omega)}{f^{m+1}}.
\end{equation}

By Stokes' theorem we have
$$ \int_{\partial \Sigma} \frac{\Delta(\omega)}{f^m} = \int_\Sigma d \left(
\frac{\Delta(\omega)}{f^m}\right). $$
Using \eqref{dDeltam3} this gives
\begin{equation}\label{intDelta1}
\int_{\partial \Sigma} \frac{\Delta(\omega)}{f^m} =m \deg(f)
\int_\Sigma \frac{\omega}{f^m} - \int_\Sigma \frac{df
\wedge \Delta(\omega)}{f^{m+1}}.
\end{equation}
\endproof

We can use this result to reformulate the parametric Feynman integrals
in terms of integrals of forms that are pullbacks to $\A^n\smallsetminus \{0\}$ of forms 
on a hypersurface complement in $\P^{n-1}$. 
For simplicity, we remove here the divergent $\Gamma$-factor from the parametric
Feynman integral and we concentrate on the residue given by the integration on 
the simplex $\Sigma$ as in \eqref{ParFeySigma} below.

\begin{prop}\label{FeynDelta}
Under the generic condition on the external momenta, the parametric Feynman integral 
\begin{equation}\label{ParFeySigma}
\bU (\Gamma)=  \int_\Sigma
\frac{\omega_n}{\Psi_\Gamma^{D/2} V_\Gamma^{n-D\ell/2}}
\end{equation}
can be computed as 
\begin{equation}\label{ParFeySigma2}
\bU(\Gamma)= \frac{1}{ C(n,D,\ell)} \left(
\int_{\partial\Sigma} \pi^*(\eta) + 
\int_{\Sigma}df \wedge \frac{\pi^*(\eta)}{f} \right),
\end{equation}
where $\pi: \A^n\smallsetminus\{0\} \to \P^{n-1}$ is the projection and $\eta$
is the form on the hypersurface complement $\cU(f)$
in $\P^{n-1}$ with 
\begin{equation}\label{etapullback}
\pi^*(\eta)=\frac{\Delta(\omega)}{f^m}, 
\end{equation}
on $\A^n$, where 
\begin{equation}\label{fnDell}
f = \left\{ \begin{array}{ll} P_\Gamma & n-\frac{D(\ell+1)}{2} \geq 0
\\[3mm]
P_\Gamma^{\frac{2n-D\ell}{2m}} \Psi_\Gamma^{\frac{D}{2m}} &
n-\frac{D(\ell+1)}{2} <0 < n-\frac{D\ell}{2} \\
& m=\gcd\{ n-D\ell/2, D/2 \} \\[3mm]
\Psi_\Gamma & n-\frac{D\ell}{2} \leq 0 ,
\end{array}\right. 
\end{equation}
with
\begin{equation}\label{mnDell}
m = \left\{ \begin{array}{ll} n- D\ell/2 & n-\frac{D(\ell+1)}{2} \geq 0
\\[2mm]
\gcd\{ n-D\ell/2, D/2 \} & n-\frac{D(\ell+1)}{2} <0 <
n-\frac{D\ell}{2} \\[2mm]
-n + D(\ell+1)/2 & n-\frac{D\ell}{2} \leq 0,
\end{array}\right.
\end{equation}
and with
\begin{equation}\label{omeganDell}
\omega= \left\{ \begin{array}{ll} \Psi_\Gamma^{n-D(\ell+1)/2} \omega_n
& n-\frac{D(\ell+1)}{2} \geq 0 \\[3mm]
\Psi_\Gamma^{n-D\ell/2} \omega_n & n-\frac{D(\ell+1)}{2} <0 <
n-\frac{D\ell}{2} \\[3mm]
P_\Gamma^{-n+D\ell/2} \omega_n & n-\frac{D\ell}{2} \leq 0,
\end{array}\right.
\end{equation}
where $\omega_n =dt_1\wedge\cdots\wedge dt_n$ on $\A^n$, with
$\Omega=\Delta(\omega_n)$ as in \eqref{volP}. The coefficient
$C(n,D,\ell)$ in \eqref{ParFeySigma2} is given by
\begin{equation}\label{CnDell}
C(n,D,\ell)= \left\{ \begin{array}{ll}
(n-D\ell/2)(\ell+1) &  n-\frac{D(\ell+1)}{2} \geq 0 \\[2mm]
(n-D\ell/2) \ell +n & n-\frac{D(\ell+1)}{2} <0 <
n-\frac{D\ell}{2} \\[2mm]
-(n-D(\ell+1)/2)\ell &  n-\frac{D\ell}{2} \leq 0.
\end{array}\right.
\end{equation}
\end{prop}

\proof Consider on $\A^n$ the form given by $\Delta(\omega)/f^m$,
with $f$, $m$, and $\omega$, respectively as in \eqref{fnDell},
\eqref{mnDell} and \eqref{omeganDell}. 
We assume the condition of Definition \ref{genericp}, \ie for a
generic choice of the external momenta the polynomials 
$P_\Gamma$ and $\Psi_\Gamma$ have
no common factor.  First notice that, since 
the polynomial $\Psi_\Gamma$ is homogeneous of degree $\ell$ and
$P_\Gamma$ is homogeneous of degree $\ell+1$, the form
$\Delta(\omega)/f^m$ is $\bG_m$ invariant on $\A^n\smallsetminus \{0\}$. Moreover, since
it is of the form $\alpha=\Delta(\omega)/f^m$, it also satisfies
$\Delta(\alpha)=0$, hence it is the pullback of a form $\eta$ on
$\cU(f)\subset \P^{n-1}$. Also notice that the domain of integration
$\Sigma \subset \A^n$ given by the simplex $\Sigma=\{ \sum_i t_i=1,\, t_i \geq
0\}$, is contained in a fundamental domain of the action of the
multiplicative group $\C^*$ on $\C^n\smallsetminus\{ 0\}$. 

Applying the result of Proposition \ref{intSigmaprop} above, we obtain
$$ \int_\Sigma \frac{dt_1 \wedge \cdots \wedge
dt_n}{\Psi_\Gamma^{D/2} V_\Gamma^{n-D\ell/2}} = \int_\Sigma
\frac{\omega}{f^m} $$
$$ = \frac{1}{m \deg(f)}( \int_{\partial\Sigma}
\frac{\Delta(\omega)}{f^m} + \int_\Sigma df \wedge
\frac{\Delta(\omega)}{f^{m+1}}) $$
$$ = C(n,D,\ell)^{-1} \left( \int_{\partial \Sigma}
\frac{\Delta(\omega_n)}{\Psi_\Gamma^{D/2} V_\Gamma^{n-D\ell/2}} 
+ \int_\Sigma df\wedge \frac{\Delta(\omega_n)}{\Psi_\Gamma^a P_\Gamma^b}
\right), $$
where $f$ is as in \eqref{fnDell} and
\begin{equation}\label{anDell}
a=\left\{\begin{array}{ll}
D(\ell+1)/2-n & n-\frac{D(\ell+1)}{2} \geq 0 \\[2mm]
\frac{D}{2} (1+\frac{1}{m}) & n-\frac{D(\ell+1)}{2} <0 <
n-\frac{D\ell}{2} \\[2mm]
-n + \frac{D(\ell+1)}{2}+1 & n-\frac{D\ell}{2} \leq 0,
\end{array}\right.
\end{equation}
\begin{equation}\label{bnDell}
b=\left\{\begin{array}{ll}
n-\frac{D\ell}{2} +1 & n-\frac{D(\ell+1)}{2} \geq 0 \\[2mm]
(n-\frac{D\ell}{2})(1+\frac{1}{m}) & n-\frac{D(\ell+1)}{2} <0 <
n-\frac{D\ell}{2} \\[2mm]
n- \frac{D\ell}{2} & n-\frac{D\ell}{2} \leq 0.
\end{array}\right.
\end{equation}
In fact, the cases of $n-\frac{D(\ell+1)}{2} \geq 0$ and
$n-\frac{D\ell}{2} \leq 0$ are clear, while in the range with
$n-\frac{D(\ell+1)}{2} <0$ and $n-\frac{D\ell}{2}>0$ we have
$$ f^{m+1}=P_\Gamma^{(n-D\ell/2)(1+\frac{1}{m})}
\Psi_\Gamma^{\frac{D}{2} (1+\frac{1}{m})}. $$
The coefficient $C(n,D,\ell)$ is given by $C(n,D,\ell)= m\deg(f)$,
with $m$ and $f$ as in \eqref{mnDell} and \eqref{fnDell}. Thus, it is
given by \eqref{CnDell}, where in the second case we use
$$ m ( (\ell+1)(n-D\ell/2)/m + D\ell/2m) = (n-D\ell/2) \ell +n, $$
for $m=\gcd\{ n-D\ell/2, D/2 \}$. 
\endproof

\section{Singularities, slicing, and Milnor fiber}\label{HodgeSec}

\subsection{Non-isolated singularities}\label{SingSec}

One problem in trying to use in our setting the techniques developed
in singularity theory (\cf \cite{AGZV}) to study mixed Hodge
structures in terms of oscillatory integrals is that the graph
hypersurfaces  
$X_\Gamma\subset \P^{n-1}$ defined by the vanishing of the polynomial
$\Psi_\Gamma(t)=\det(M_\Gamma(t))$ usually have non-isolated
singularities. This can be seen easily by the following observation. 

\begin{lem}\label{singXGamma}
Let $\Gamma$ be a graph with $\deg \Psi_\Gamma > 2$.
The singular locus of $X_\Gamma$ is given by the intersection of cones
over the hypersurfaces $X_{\Gamma_e}$, for $e\in E(\Gamma)$, where
$\Gamma_e$ is the graph obtained by removing the edge $e$ of
$\Gamma$. The cones $C(X_{\Gamma_e})$ do not intersect transversely. 
\end{lem}

\proof First observe that, since $X_\Gamma$ is defined by a
homogeneous equation $\Psi_\Gamma(t)=0$, with $\Psi_\Gamma$ a
polynomial of degree $m$, the Euler formula $m \Psi_\Gamma(t) =\sum_e
t_e\frac{\partial}{\partial t_e} \Psi_\Gamma(t)$ implies that $\cap_e
Z(\partial_e \Psi_\Gamma) \subset X_\Gamma$, where $Z(\partial_e \Psi_\Gamma)$ is
the zero locus of the $t_e$-derivative. Thus, the singular locus of
$X_\Gamma$ is just given by the equations $\partial_e \Psi_\Gamma
=0$. The variables $t_e$ appear in the polynomial $\Psi_\Gamma(t)$
only with degree zero or one,  hence the polynomial $\partial_e
\Psi_\Gamma$ consists of only those monomials of $\Psi_\Gamma$ that
contain the variable $t_e$, where one sets $t_e=1$. The resulting
polynomial is therefore of the form $\Psi_{\Gamma_e}$, where
$\Gamma_e$ is the graph obtained from $\Gamma$ by removing the edge
$e$.  In fact, one can see in terms of spanning trees that, if $T$ is
a spanning tree containing the edge $e$ then $T\smallsetminus e$ is no
longer a spanning tree of $\Gamma_e$, so the corresponding terms
disappear in passing from $\Psi_\Gamma$ to $\Psi_{\Gamma_e}$, while if
$T$ is a spanning tree of $\Gamma$ which does not contain $e$, then
$T$ is still a spanning tree of $\Gamma_e$ and the corresponding
monomial $m_T$ of $\Psi_{\Gamma_e}$ is the same as the monomial $m_T$
in $\Psi_\Gamma$ without the variable $t_e$.  Thus, the zero locus
$Z(\Psi_{\Gamma_e})\subset \P^{n-1}$ is a cone $C(X_{\Gamma_e})$ over
the graph hypersurface $X_{\Gamma_e}\subset \P^{n-2}$ with vertex at
the coordinate point $v_e=(0,\ldots,0,1,0,\ldots 0)$ with $t_e=1$. To
see that these cones do not intersect transversely, notice that, in
the case where $\deg \Psi_\Gamma >2$, given any two $C(X_{\Gamma_e})$
and $C(X_{\Gamma_{e'}})$ the vertex of one cone is contained in the
graph hypersurface spanning the other cone. 
\endproof

The work of Bergbauer--Rej \cite{BerRej} gives a more detailed
analysis of the singular locus of the graph hypersurfaces, using a
formula for the Kirchhoff polynomials under insertion of subgraphs at
vertices. 

\medskip

\subsection{Projective Radon transform}\label{RadonSec}

Among various techniques introduced for the study of non-isolated
singularities, a common procedure consists of cutting the ambient space
with linear spaces of dimension complementary to that of the singular
locus of the hypersurface (\cf \eg \cite{Teiss}). 
In this case, the restriction of the
function defining the hypersurface to these linear spaces defines 
hypersurfaces with isolated singularities, to which the usual
invariants and constructions for isolated singularities can be
applied. 

\smallskip

One finds that, in typical cases, the graph hypersurfaces have singular
locus of codimension one, which means that the slicing is given by planes
$\P^2$ intersecting the hypersurface into a curve with isolated singular 
points. When the singular locus is of codimension two in the hypersurface, 
the slicing is given by 3-dimensional spaces cutting the hypersurface into a family of
surfaces in $\P^3$ with isolated singularities. 

\smallskip

In our setting, we are interested in computing integrals of the 
form \eqref{ParFeySigma}. From this point of view, the procedure of
restricting the function defining the hypersurface to linear spaces of
a fixed dimension correspond to an integral transform analogous to a
Radon transform in projective space (\cf \cite{GGG}).

\smallskip

We recall the basic setting for integral transforms on projective
spaces (\cf \S II of \cite{GGG}). 
On any $k$-dimensional subspace $\A^k\subset \A^n$ there is a
unique (up to a multiplicative constant) $(k-1)$-form that is
invariant under the action of $\SL_k$. It is given as in \eqref{volP}
by the expression
\begin{equation}\label{Omegak}
 \Omega_k= \sum_{i=1}^k (-1)^{i+1} t_i \, dt_1 \wedge \cdots \wedge \widehat{dt_i}
\wedge \cdots \wedge dt_k. 
\end{equation}
The form \eqref{Omegak} is homogeneous of degree $k$.
Suppose given a function $f$ on $\A^n$ which
satisfies the homogeneity condition
\begin{equation}\label{homogk}
f(\lambda t) = \lambda^{-k} f(t), \ \ \ \forall t\in \A^n, \, \lambda\in \bG_m.
\end{equation}
Then the integrand $f \Omega_k$ is well
defined on the corresponding projective space $\P^{k-1}\subset
\P^{n-1}$ and one defines the integral as integrating 
on a fundamental domain in $\A^k\smallsetminus \{0\}$, \ie on a surface that intersect
each line from the origin once. 

\smallskip

Suppose given dual vectors $\xi_i \in (\A^n)^\prime$, for
$i=1,\ldots,n-k$. These define a $k$-dimensional linear subspace
$\Pi=\Pi_\xi \subset \A^n$ by the vanishing
\begin{equation}\label{Pixi}
\Pi_\xi =\{ t\in \A^n \,|\, \langle \xi_i, t \rangle =0, \,
i=1,\ldots,n-k \}.
\end{equation}
Given a choice of a subspace $\Pi_\xi$, there exists a $(k-1)$-form
$\Omega_\xi$ on $\A^n$ satisfying
\begin{equation}\label{Omegaxi}
\langle \xi_1,dt \rangle \wedge \cdots \wedge \langle \xi_{n-k},
dt \rangle \wedge \Omega_\xi = \Omega_n,
\end{equation}
with $\Omega_n$ the $(n-1)$-form of \eqref{volP}, \cf \eqref{Omegak}. The form
$\Omega_\xi$ is not uniquely defined on $\A^n$, but its restriction to
$\Pi_\xi$ is uniquely defined by \eqref{Omegaxi}. Then, given a
function $f$ on $\A^n$ with the homogeneity condition \eqref{homogk},
one can consider the integrand $f \Omega_\xi$ and define its integral
on the projective space $\pi(\Pi_\xi)\subset \P^{n-1}$ as above. This
defines the integral transform, that is, the $(k-1)$-dimensional projective Radon
transform (\S II of \cite{GGG}) as
\begin{equation}\label{kRadon}
\cF_k(f)(\xi)= \int_{\pi(\Pi_\xi)} f(t) \, \Omega_\xi(t) =
\int_{\P^{n-1}} f(t) \prod_{i=1}^{n-k} \delta(\langle \xi_i,t \rangle)
\, \Omega_\xi(t).
\end{equation}

\smallskip

For our purposes, it is convenient to consider also the following
variant of the Radon transform \eqref{kRadon}.

\begin{defn}\label{SigmaRadon}
Let $\Sigma \subset \A^n$ be a compact region that is contained in a
fundamental domain of the action of $\bG_m$ on $\A^n\smallsetminus\{0\}$. The
partial $(k-1)$-dimensional projective Radon transform is given by the expression
\begin{equation}\label{kRadonSigma}
\cF_{\Sigma,k}(f)(\xi)= \int_{\Sigma \cap \pi(\Pi_\xi)} f(t) \, \Omega_\xi(t) =
\int_{\Sigma\cap \pi(\Pi_\xi)} f(t) \prod_{i=1}^{n-k} \delta(\langle \xi_i,t \rangle)
\, \Omega_\xi(t),
\end{equation}
where one identifies $\Sigma$ with its image $\pi(\Sigma)\subset \P^{n-1}$. 
\end{defn}

\medskip

Let us now return to the parametric Feynman integrals we are
considering.

\begin{prop}\label{FeynRadon}
The Feynman integral \eqref{expint} can be reformulated as
\begin{equation}\label{expintRad}
U(\Gamma)= \frac{\Gamma(k-\frac{D\ell}{2})}{ (4\pi)^{\ell D/2}} 
\int \cF_{\Sigma,k}(f_\Gamma)(\xi)\,  \langle \xi, dt \rangle,
\end{equation}
where $\xi$ is an $(n-k)$-frame in $\A^n$ and $\cF_{\Sigma,k}(f)$ is
the Radon transform, with $\Sigma$ the simplex $\sum_i t_i=1$,
$t_i\geq 0$, and with
\begin{equation}\label{ftxi} 
f_\Gamma(t)= \frac{V_\Gamma(t,p)^{-k+D\ell/2}}{\Psi_\Gamma(t)^{D/2}} .
\end{equation}
\end{prop}

\proof
Consider first the form \eqref{expint} of the Feynman integral, which
we write equivalently as
\begin{equation}\label{expintA}
U(\Gamma)= (4\pi)^{-\ell D/2}\, \int_{\A^n} \chi_+(t) 
\frac{e^{-V_\Gamma(t,p)}}{\Psi_\Gamma(t)^{D/2}} \,\, dt_1 \cdots dt_n,    
\end{equation}
where $\chi_+(t)$ is the characteristic function of the domain
$\R^n_+$. 

Given a choice of an $(n-k)$-frame $\xi$, we
can then write the Feynman integrals in the form
\begin{equation}\label{expintPi}
U(\Gamma)=(4\pi)^{-\ell D/2}\, \int 
\left( \int_{\Pi_\xi} \chi_+(t)
\frac{e^{-V_\Gamma(t,p)}}{\Psi_\Gamma(t)^{D/2}} \omega_\xi
\right)\, \langle \xi, dt \rangle ,
\end{equation}
where $\langle \xi, dt \rangle$ is a shorthand notation for
$$ \langle \xi, dt \rangle = \langle \xi_1,dt \rangle \wedge \cdots
\wedge \langle\xi_{n-k}, dt \rangle $$
and $\omega_\xi$ satisfies
\begin{equation}\label{omegaxi}
 \langle \xi, dt \rangle \wedge \omega_\xi =\omega_n
= dt_1\wedge\cdots\wedge dt_n . 
\end{equation}

We then apply the same procedure as in \eqref{expint2} and
\eqref{Vdlambda} to the integral on $\Pi_\xi$ and write it in the form
\begin{equation}\label{expintPiV}
\int_{\Pi_\xi} \chi_+(t)
\frac{e^{-V_\Gamma(t,p)}}{\Psi_\Gamma(t)^{D/2}} \omega_\xi(t)  =
\Gamma(k-\frac{D\ell}{2}) \int_{\Pi_\xi} \delta(1-\sum_i t_i)\,
\frac{\omega_\xi(t)}{\Psi_\Gamma(t)^{D/2}V_\Gamma(t,p)^{k-D\ell/2}}.
\end{equation}

The function $f_\Gamma(t)$ of \eqref{ftxi}
satisfies the scaling property \eqref{homogk} and the integrand
$$ 
\frac{\omega_\xi(t)}{\Psi_\Gamma(t)^{D/2}V_\Gamma(t,p)^{k-D\ell/2}}
$$
is therefore $\bG_m$-invariant, since the form $\omega_\xi$ is homogeneous of
degree $k$. Moreover, the domain
$\Sigma$ of integration is contained in a 
fundamental domain for the action of $\bG_m$. Thus, 
we can reformulate the integral \eqref{expintPiV} in 
projective space, in terms of Radon transform as
\begin{equation}\label{expintPiF}
\Gamma(k-\frac{D\ell}{2})\,(4\pi)^{-\ell D/2}\, \int
\cF_{\Sigma,k}(f_\Gamma)(\xi)\,  \langle \xi, dt \rangle ,
\end{equation}
where $\cF_{\Sigma,k}(f_\Gamma)$ is the Radon transform over the
simplex $\Sigma$, as in Definition \ref{SigmaRadon}. 
\endproof

\smallskip

In the following, we will then consider integrals of the form
\begin{equation}\label{UGammaxi}
\bU(\Gamma)_\xi = \cF_{\Sigma,k}(f_\Gamma)(\xi) = \int_{\Pi_\xi} \delta(1-\sum_i t_i)\,
\frac{\omega_\xi(t)}{\Psi_\Gamma(t)^{D/2}V_\Gamma(t,p)^{k-D\ell/2}} 
\end{equation}
$$ = \int_{\Sigma_\xi}
\frac{\omega_\xi(t)}{\Psi_\Gamma(t)^{D/2}V_\Gamma(t,p)^{k-D\ell/2}} $$
as well as their dimensional regularizations
\begin{equation}\label{UGammaxiDR}
\bU(\Gamma)_\xi(z) =\int_{\Sigma_\xi}
\frac{\omega_\xi(t)}{\Psi_\Gamma(t)^{(D+z)/2}
V_\Gamma(t,p)^{k-(D+z)\ell/2}},
\end{equation}
where $\Pi_\xi$ is a generic linear subspace of dimension equal to the
codimension of the singular locus of the hypersurface $X_\Gamma\cup
Y_\Gamma$.

\medskip

\subsection{The polar filtration}\label{PolarSec}

As we recalled already in \S \ref{IntProjSec} above (\cf \cite{Dimca})
algebraic differential forms $\omega \in \Omega^k(\cD(f))$
on a hypersurface complement can always be written in the form $\omega
=\eta/f^m$ as in \eqref{formfm}, for some $m\in \N$ and some
$\eta\in \Omega^k_{m\deg(f)}$. The minimal $m$ such that $\omega$ can
be written in the form $\omega=\eta/f^m$ is called the order of pole 
of $\omega$ along the hypersurface $X$ and is denoted by $\ord_X(\omega)$.
The order of pole induces a filtration, called the {\em polar
filtration}, on the de Rham complex of differential forms on the 
hypersurface complement. One denotes by $P^r \Omega^k_{\P^n}
\subset\Omega^k_{\P^n}$ the subspace of forms of order 
${\rm ord}_X(\omega)\leq k-r+1$, if $k-r+1\geq 0$, 
or $P^r \Omega^k=0$ for $k-r+1< 0$. The polar filtration
$P^\bullet$ is related to the Hodge filtration $F^\bullet$ by 
$P^r \Omega^m \supset F^r \Omega^m$, by a result of \cite{DeDi}.

\medskip

\begin{prop}\label{Prkxi}
Under the generic condition on the external momenta, the forms 
\begin{equation}\label{xiformsDell}
\frac{\Omega_\xi}{\Psi_\Gamma^{D/2}
V_\Gamma^{k-D\ell/2}}
\end{equation}
span subspaces $P^{r,k}_\xi$ of the polar filtration $P^r
\Omega^{k-1}_{\P^{n-1}}$ of a hypersurface complement $\cU(f)\subset
\P^{n-1}$, where
\begin{equation}\label{frkDell}
f=\left\{ \begin{array}{ll} P_\Gamma & k-D(\ell+1)/2 \geq 0 \\[3mm]
P_\Gamma^{(k-D\ell/2)/m} \Psi_\Gamma^{D/(2m)} & k-D(\ell+1)/2< 0 <
k-D\ell/2, \\ & m=\gcd \{ k-D\ell/2, D/2 \} \\[3mm]
\Psi_\Gamma & k-D\ell/2 \leq 0,
\end{array} \right.
\end{equation}
and for the index $r$ of the filtration in the range
\begin{equation}\label{ranger}
\left\{\begin{array}{ll}
r \leq D\ell/2 & k-D(\ell+1)/2 \geq 0 \\[2mm]
r\leq k-\gcd \{ k-D\ell/2, D/2 \} & k-D(\ell+1)/2< 0 <
k-D\ell/2 \\[2mm]
r\leq 2k -D(\ell+1)/2 & k-D\ell/2 \leq 0.
\end{array}\right.
\end{equation}
\end{prop}

\proof We are assuming that $P_\Gamma$ and $\Psi_\Gamma$ have
no common factor, for generic external momenta.
Consider first the case where $k-D\ell/2 \geq 0$. This is further
divided into two cases: the case where also $k-D(\ell+1)/2 \geq 0$ and
the case where $k-D(\ell+1)/2 <0$. In the first case, the form
\eqref{xiformsDell} can be written, using \eqref{ratioVGamma}, as
\begin{equation}\label{xiformsDell1}
\frac{\Delta(\alpha)}{f^m}=\frac{\Psi_\Gamma^{k-D(\ell+1)/2}\Omega_\xi}{P_\Gamma^{k-D\ell/2}},
\end{equation}
where
\begin{equation}\label{alpham1}
 \alpha = \Psi_\Gamma^{k-D(\ell+1)/2} \omega_\xi \ \ \ \text{ and }
\ \ \  f = P_\Gamma, \ \ \ \text{ with } \ \ \  m=k-D\ell/2. 
\end{equation}
Thus, in this case we considered the polar filtration for differential
forms on the complement of the projective hypersurface $Y_\Gamma$
of degree $\ell+1$ defined by $P_\Gamma =0$.
The forms \eqref{xiformsDell1}, for a generic choice of the
$(n-k)$-frame $\xi$, and for varying external momenta $p$, span
a subspace $P^{r,k}_\xi$ of the polar filtration $P^r
\Omega^{k-1}_{\P^{n-1}}$, for all $r \leq D\ell/2$. Notice that $r\leq
D\ell/2$ also implies $r\leq k$ so that one remains within the
nontrivial range $k-r\geq 0$ of the filtration. 

In the case where we still have $k-D\ell/2 \geq 0$ but $k-D(\ell+1)/2
<0$, we let
$$ m = \gcd \{ k-D\ell/2, D/2 \}, $$
so that $k-D\ell/2=n_1 m$ and $D/2= n_2 m$. We then write
\eqref{xiformsDell} in the form
\begin{equation}\label{xiformsDell2}
\frac{\Delta(\alpha)}{f^m}=\frac{\Psi_\Gamma^{k-D\ell/2}\Omega_\xi}
{P_\Gamma^{k-D\ell/2}\Psi_\Gamma^{D/2}},
\end{equation}
with
\begin{equation}\label{alpham2}
\alpha = \Psi_\Gamma^{k-D\ell/2}\omega_\xi, \ \ \ \text{ and } \ \ \
f = P_\Gamma^{n_1} \Psi_\Gamma^{n_2} \ \ \ \text{ and } \ \ \ m=\gcd \{ k-D\ell/2, D/2 \}.
\end{equation}
In this case, we consider the polar filtration associated to the
complement of the projective hypersurface defined by the equation
$P_\Gamma^{n_1} \Psi_\Gamma^{n_2}=0$. For a generic choice of the
$(n-k)$-frame $\xi$, and for varying external momenta $p$,
we obtain in this case a subspace $P^{r,k}_\xi$ of the polar filtration $P^r
\Omega^{k-1}_{\P^{n-1}}$, for all $r\leq k-\gcd \{ k-D\ell/2, D/2 \}$.

The remaining case is when $k-D\ell/2 <0$, so that also $k-D(\ell+1)/2
<0$. In this case, we write \eqref{xiformsDell} in the form
\begin{equation}\label{xiformsDell3}
\frac{\Delta(\alpha)}{f^m}= \frac{P_\Gamma^{-k+D\ell/2}\Omega_\xi}{\Psi_\Gamma^{-k+D(\ell+1)/2}},
\end{equation}
where
\begin{equation}\label{alpham3}
\alpha = P_\Gamma^{-k+D\ell/2}\omega_\xi, \ \ \ \text{ and } \ \ \
f = \Psi_\Gamma \ \ \ \text{ and } \ \ \ m=-k+D(\ell+1)/2.
\end{equation}
We are considering here the polar filtration on forms on the
complement of the hypersurface $X_\Gamma$ defined by $\Psi_\Gamma=0$. 
We then obtain, for generic $\xi$ and varying $p$, a subspace of $P^{r,k}_\xi$ of the
filtration $P^r\Omega^{k-1}_{\P^{n-1}}$, for all $r\leq 2k -D(\ell+1)/2$.
\endproof

\medskip

\subsection{Milnor fiber}\label{MilFibSec}

Suppose then that $k={\rm codim}\, {\rm Sing}(X)$, where ${\rm
Sing}(X)$ is the singular locus of the hypersurface $X=\{ f=0 \}$,
with $f$ as in Proposition \ref{Prkxi} above. In this case, for
generic $\xi$, the linear space $\Pi_\xi$ cuts the singular locus
${\rm Sing}(X)$ transversely and the restriction $X_\xi=X\cap \Pi_\xi$
has isolated singularities. 

\smallskip

Recall that, in the case of isolated singularities, there is an
isomorphism between the cohomology of the Milnor fiber $F_\xi$ of
$X_\xi$ and the total cohomology of the Koszul--deRham complex of
forms \eqref{formfm} with the total differential $d_f \omega = 
f\, d\omega - m\, df \wedge \omega$ as above. 
The explict isomorphism is given by the Poincar\'e residue map 
and can be written in the form 
\begin{equation}\label{ResjF}
[\omega] \mapsto [j^*\Delta(\omega_\xi)],
\end{equation}
where $j: F_\xi\hookrightarrow \Pi_\xi$ is the inclusion of the 
Milnor fiber in the ambient space (see \cite{Dimca}, \S 6).

Let $M(f)$ be the Milnor algebra of $f$, \ie the quotient of the 
polynomial ring in the coordinates of the ambient projective space 
by the ideal of the derivatives of $f$. When $f$ has isolated 
singularities, the Milnor algebra is finite dimensional. One denotes
by $M(f)_m$ the homogeneous component of degree $m$ of $M(f)$.

It then follows from the identification \eqref{ResjF} above (\cite{Dimca},\S
6.2) that, in the case of isolated singularities, a basis for the 
cohomology $H^r(F_\xi)$ of the Milnor fiber, with $r=\dim \Pi_\xi -1$ 
is given by elements of the form
\begin{equation}\label{gensMilnor}
\omega_\alpha= \frac{t^\alpha \Delta(\omega_\xi)}{f^m}, \ \ \ \ 
\text{ with } \ \ \  t^\alpha \in M(f)_{m\deg(f)-k},
\end{equation}
where $f$ is the restriction to $\Pi_\xi$ of the function of
\eqref{frkDell}. 
We then have the following consequence of Proposition \ref{Prkxi}.

\begin{cor}\label{MilFibH}
For a generic $(n-k)$-frame $\xi$ with $n-k=\dim {\rm Sing}(X)$, with
$X$ the hypersurface of Proposition \ref{Prkxi}, and for a fixed
generic choice of the external momenta $p$ under the assumption of
Definition \ref{genericp},  
the Feynman integrand \eqref{xiformsDell} of \eqref{UGammaxi} defines
a cohomology class in $H^r(F_\xi)$, with $r=\dim \Pi_\xi-1$ and
$F_\xi\subset \Pi_\xi$ the Milnor fiber of the hypersurface with
isolated singularities $X_\xi=X\cap \Pi_\xi \subset \Pi_\xi$. 
\end{cor}

\proof By Proposition \ref{Prkxi}, the form \eqref{xiformsDell} can be
written as
\begin{equation}\label{hDeltafm}
\frac{h\Delta(\omega_\xi)}{f^m},
\end{equation}
where $f$ is as in \eqref{frkDell}, and $h$ is a polynomial of the form
\begin{equation}\label{hkDell}
h=\left\{ \begin{array}{ll}
\Psi_\Gamma^{k-D(\ell+1)/2}    & k-D(\ell+1)/2 \geq 0 \\[2mm]
\Psi_\Gamma^{k-D\ell/2}    & k-D(\ell+1)/2< 0 < k-D\ell/2 \\[2mm]
P_\Gamma^{-k+D\ell/2}    & k-D\ell/2 \leq 0.
\end{array}\right.
\end{equation}
Let $\cI_\xi$ denote the ideal of derivatives of the restriction $f|_{\Pi_\xi}$ of
$f$ to $\Pi_\xi$. Then let
\begin{equation}\label{hxi}
h_\xi = h \mod \cI_\xi.
\end{equation}
For a fixed generic choice of the external momenta, this defines an
element in the Milnor algebra $M(f|_{\Pi_\xi})$, which 
lies in the homogeneous component $M(f|_{\Pi_\xi})_{m\deg(f)-k}$, for
\begin{equation}\label{mkDell}
 m=\left\{ \begin{array}{ll} k-D\ell/2 & k-D(\ell+1)/2 \geq 0
\\[2mm] \gcd\{ k-D\ell/2, D/2\} & k-D(\ell+1)/2< 0< k-D\ell/2 \\[2mm]
-k + D(\ell+1)/2 & k-D\ell/2 \leq 0.
\end{array}\right. 
\end{equation}
Thus, the form \eqref{hDeltafm} defines a class in the cohomology
$H^r(F_\xi)$ with $r=\dim \Pi_\xi-1$. 
\endproof

\medskip

\subsection{The Feynman integral: slicing}\label{SliceSec}

As in Proposition \ref{FeynDelta}, we can reformulate the integral
\eqref{UGammaxi} in terms of integrals of pullbacks of forms on a
hypersurface complement in projective space, using the explicit
description of Proposition \ref{Prkxi} above.

\begin{prop}\label{xiFeynDelta}
The integral \eqref{UGammaxi} can be computed in the form
\begin{equation}\label{ParFeySigmaxi}
\bU(\Gamma)_\xi =\frac{1}{C(k,D,\ell)} \left(
\int_{\partial\Sigma \cap \Pi_\xi} \pi^*(\eta_\xi) + 
\int_{\Sigma \cap \Pi_\xi} df|_{\Pi_\xi}\wedge \frac{\pi^*(\eta_\xi)}{f|_{\Pi_\xi}} \right),
\end{equation}
where $\pi: \A^n\smallsetminus\{0\} \to \P^{n-1}$ is the projection and $\eta_\xi$
satisfies
\begin{equation}\label{etapullxi}
\pi^*(\eta_\xi)=\frac{h|_{\Pi_\xi} \Omega_\xi}{(f|_{\Pi_\xi})^m} 
\end{equation}
on $\A^n$, where $\Omega_\xi$ is given by \eqref{Omegaxi} and $f$,
$m$ and $h$ are as in Proposition \ref{Prkxi} and Corollary
\ref{MilFibH}. The coefficient $C(k,D,\ell)$ is given as in
\eqref{CnDell}.
\end{prop}

\proof As in the case of Proposition \ref{FeynDelta}, the result
follows by applying Proposition \ref{intSigmaprop} and Proposition \ref{Prkxi},
together with the fact that $\Omega_\xi = \Delta(\omega_\xi)$, 
which can be seen by writing
$$ \Omega_\xi = \prod_{i=1}^{n-k} \delta(\langle \xi_i, t\rangle) \,
\Omega_n . $$
The coefficient $C(k,D,\ell)$ is given by $C(k,D,\ell)=m\deg(f)$, 
with $m$ and $f$ as in \eqref{mkDell} and \eqref{frkDell}.
\endproof

\section{Oscillatory integrals: Leray and Dimensional Regularizations}\label{OscSec}

A well known method for studying integrals of holomorphic forms on
vanishing cycles of a singularity and to relate these to mixed Hodge
structures is via oscillatory integrals and their asymptotic expansion
(see \cite{AGLV} and Vol.II of \cite{AGZV}). Our main result in this
section will be to show that the dimensionally regularized parametric 
Feynman integrals can be related to the Mellin transform of a
Gelfand--Leray form, whose Fourier transform is the oscillatory
integral usually considered in the context of singularity theory.

\medskip

\subsection{Oscillatory integrals and the Gelfand--Leray forms}\label{GLSec}

We recall briefly some results on oscillatory integrals and their
asymptotic expansion. We refer the reader to \S 2, Vol.II of
\cite{AGZV} for more details. 
In general, an {\em oscillatory integral} is an expression of the form
\begin{equation}\label{intfphi}
I(\alpha)= \int_{\R^n} e^{i\alpha f(x)} \phi(x) \, dx_1 \cdots dx_n,
\end{equation}
where $f:\R^n \to \R$ and $\phi: \R^n \to \R$ are smooth functions and
$\alpha \in \R^*_+$  
is a real parameter.  It is well known that, if the phase $f(x)$ is an
analytic function in a neighborhood  
of a critical point $x_0$, then \eqref{intfphi} has an asymptotic
development for $\alpha\to \infty$  
given by a series
\begin{equation}\label{asynserf}
I(\alpha) \sim e^{i\alpha f(x_0)} \sum_u \sum_{k=0}^{n-1}
a_{k,u}(\phi) \, \alpha^u (\log \alpha)^k , 
\end{equation}
where $u$ runs over a finite set of arithmetic progressions of
negative rational numbers depending only on the phase $f(x)$, and the
$a_{k,u}$ are distributions supported on the critical points of the
phase, \cf \S 2.6.1, Vol.II of \cite{AGZV}.  

\medskip

It is also well known that the integral \eqref{intfphi} can be
reformulated in terms of one-dimensional integrals using the
Gelfand--Leray form 
\begin{equation}\label{intGLform}
I(\alpha)= \int_\R  e^{i\alpha t} \left( \int_{X_t(\R)} \phi(x) \omega_f(x,t) \right)  \,\, dt 
\end{equation}
where $X_t(\R)\subset \R^n$ is the level set $X_t(\R)=\{x\in\R^n\,:\,
f(x)=t\}$ and $\omega_f(x,t)$ is the Gelfand--Leray form, that is, the
unique $(n-1)$-form on the level hypersurface $X_t$ with the property
that 
\begin{equation}\label{GLform}
df \wedge \omega_f(x,t) = dx_1\wedge\cdots \wedge dx_n. 
\end{equation}
Notice that, as in the case of the forms \eqref{Omegaxi}, there is
some choice of an $(n-1)$-form satisfying \eqref{GLform}, but the 
restriction to $X_t$ is unique so that the Gelfand--Leray form on
$X_t$ is well defined.  
Notice also that, up to throwing away a set of measure zero, we can
assume here that the integration is over the values $t\in \R$ such
that the level set $X_t$ is a smooth hypersurface.  

The Gelfand--Leray form $\omega_f(x,t)$ is often written in the notation
\begin{equation}\label{GLformdf}
\omega_f(x,t)= \frac{dx_1\wedge\cdots \wedge dx_n}{df}.
\end{equation}
It is given by the Poincar\'e residue
\begin{equation}\label{PoincRes}
\frac{\omega}{df} = \Res_{\epsilon=0}\, \frac{\omega}{f-\epsilon}.
\end{equation}

The Gelfand--Leray function is the associated function
\begin{equation}\label{GLfunct}
J(t):=  \int_{L_t} \phi(x) \omega_f(x,t) .
\end{equation}
For more details, see \S 2.6 and \S 2.7, Vol.II \cite{AGZV}.

\medskip

We recall here a property of the Gelfand--Leray forms that will be
useful in the following, where we consider complex hypersurfaces
$X\subset \A^n=\C^n$, with defining 
polynomial equation $f=0$ and the hypersurface complement
$\cD(f)\subset \A^n$, such that the restriction of $f$ to the interior of
the domain of integration $\Sigma \subset \A^n$ takes values in $\R^*_+$. 

Recall that the Leray coboundary of a $k$-chain $\sigma$ in $X$ is a
$(k+1)$-chain in $\cD(f)$ obtained by considering a tubular
neighborhood of $X$ in $\A^n$, in the following way. Since $X$ is a hypersurface, the
boundary of its tubular neighborhood is a circle bundle over $X$. One
considers the preimage of $\sigma$ under the projection map as a chain
in $\cD(f)$. We denote the resulting chain by $\cL(\sigma)$. It is
called the Leray coboundary of $\sigma$ (see \cite{AGZV} p.282). 
The Leray coboundary $\cL(\sigma)$ is a cycle if $\sigma$ is a cycle, 
and if one changes $\sigma$ by a boundary then $\cL(\sigma)$ also
changes by a boundary. 

\begin{lem}\label{dfintLem}
Let $\sigma_\epsilon$ be a $k$-chain in $X_\epsilon=\{ t\in \A^n |
f(t)=\epsilon\}$ and let $\cL(\sigma_\epsilon)$ be its Leray coboundary
in $\cD(f-\epsilon)$. Then, for a form $\alpha\in \Omega^k$ that
admits a Gelfand--Leray form, one has
\begin{equation}\label{dfint}
\frac{1}{2\pi i} \int_{\cL(\sigma(\epsilon))} df \wedge
\frac{\alpha}{f-\epsilon} = \int_{\sigma(\epsilon)} \alpha, 
\end{equation}
where
\begin{equation}\label{depsilon}
\frac{d}{d\epsilon} \int_{\sigma(\epsilon)} \alpha =
\int_{\sigma(\epsilon)} \frac{d\alpha}{df} -
\int_{\partial\sigma(\epsilon)} \frac{\alpha}{df}.
\end{equation}
\end{lem}

\proof First let us show that if $\alpha$ has a Gelfand--Leray form
then $d\alpha$ also does. We have a form $\alpha/df$ such that
$$ df \wedge \frac{\alpha}{df} = \alpha. $$
Its differential gives
$$ d\alpha = d\left( df \wedge \frac{\alpha}{df} \right) = -df \wedge
d\left(\frac{\alpha}{df} \right). $$
Thus, the form 
$$ \frac{d\alpha}{df} = -d\left(\frac{\alpha}{df} \right)  $$
is a Gelfand--Leray form for $d\alpha$. 

Then we proceed to prove the first statement. One can write
$$ \frac{1}{2\pi i} \int_{\cL(\sigma(\epsilon))} df \wedge
\frac{\alpha}{f-\epsilon} = \frac{1}{2\pi i} \int_{\gamma} \left(
\int_{\sigma(s)} \alpha \right) \, \frac{ds}{s-\epsilon} , $$
where $\gamma\cong S^1$ is the boundary of a small disk centered at
$\epsilon \in \C$. This can then be written as
$$ = \frac{1}{2\pi i} \int_{\gamma} 
\int_{\sigma(\epsilon)} \alpha  \, \frac{ds}{s-\epsilon} + \left(
\frac{1}{2\pi i} \int_{\gamma} 
\int_{\sigma(s)} \alpha  \, \frac{ds}{s-\epsilon} 
- \frac{1}{2\pi i} \int_{\gamma}\int_{\sigma(\epsilon)} \alpha \,
\frac{ds}{s-\epsilon} \right). $$
The last term can be made arbitrarily small, so one gets
\eqref{dfint}. To obtain \eqref{depsilon} notice that
$$ \frac{1}{2\pi i} \frac{d}{d\epsilon}  \int_{\cL(\sigma(\epsilon))} df \wedge
\frac{\alpha}{f-\epsilon} = \frac{1}{2\pi i}
\int_{\cL(\sigma(\epsilon))} df\wedge \frac{\alpha}{(f-\epsilon)^2}. $$
One then uses
$$ d\left( \frac{\alpha}{f-\epsilon} \right) =
\frac{d\alpha}{f-\epsilon} - \frac{\alpha}{(f-\epsilon)^2} $$
to rewrite the above as
$$ \frac{1}{2\pi i} \left( \int_{\cL(\sigma(\epsilon))}
\frac{d\alpha}{f-\epsilon} - \int_{\cL(\sigma(\epsilon))} 
d\left( \frac{\alpha}{f-\epsilon} \right) \right) $$
$$ = \frac{1}{2\pi i} \int_{\cL(\sigma(\epsilon))} \frac{df \wedge
\frac{d\alpha}{df}}{f-\epsilon} - \frac{1}{2\pi i}    
\int_{\cL(\partial \sigma(\epsilon))} \frac{\alpha}{f-\epsilon} $$
$$ = \frac{1}{2\pi i} \int_{\cL(\sigma(\epsilon))} \frac{df \wedge
\frac{d\alpha}{df}}{f-\epsilon} - \frac{1}{2\pi i}    
\int_{\cL(\partial \sigma(\epsilon))} \frac{df \wedge \frac{\alpha}{df}}{f-\epsilon}, $$
where $d\alpha/df$ is a Gelfand--Leray form such that
$$ df \wedge \frac{d\alpha}{df} = d\alpha, $$
and $\alpha/df$ is a Gelfand--Leray form with the property that
$$ df\wedge \frac{\alpha}{df} = \alpha. $$
This then gives by \eqref{dfint}
$$ \frac{d}{d\epsilon} \int_{\sigma(\epsilon)} \alpha =
\int_{\sigma(\epsilon)} \frac{d\alpha}{df} -
\int_{\partial\sigma(\epsilon)} \frac{\alpha}{df}. $$
This completes the proof.
\endproof

\medskip

\subsection{Leray coboundary regularization and subtraction}\label{LerRegSec}

The formulation \eqref{ParFeySigma2} of the parametric Feynman
integrals, in the form of Proposition \ref{FeynDelta}, suggests a
regularization procedure different from Dimensional Regularization,
but with the similar effect of replacing a divergent integral with a
meromorphic function to which the ``minimal subtraction'' procedure
can be applied to remove the polar part and extract a finite value.

\smallskip

Since the singularities arise where the domain of integration $\Sigma$
meets the hypersurface $X=\{ f=0 \}$, with $f$ as in \eqref{fnDell},
we can concentrate on only the part of the integral that is supported
near this intersection.  

\smallskip 

Let $D_\epsilon(X)$ denote a neighborhood of the hypersurface $X$ in
$\P^{n-1}$, given by level sets
\begin{equation}\label{DepsilonX}
D_\epsilon(X) = \cup_{s\in \Delta^*_\epsilon} X_s,
\end{equation}
where $X_s=\{ t| f(t)=s\}$ and $\Delta^*_\epsilon\subset \C^*$ is a
small punctured disk of radius $\epsilon >0$. The boundary 
$\partial D_\epsilon(X)$ is given by
\begin{equation}\label{partialDepsilonX}
\partial D_\epsilon(X) = \cup_{s\in \partial\Delta^*_\epsilon} X_s.
\end{equation}
It is a circle bundle over the generic fiber $X_\epsilon$, with
projection $\pi_\epsilon: \partial D_\epsilon(X)\to X_\epsilon$.
Given a domain of integration $\Sigma$, we consider the intersection 
$\Sigma \cap D_\epsilon(X)$. This is the region that contains the
locus $\Sigma \cap X$ where the divergence in the Feynman integral 
can occur. We let $\cL_\epsilon(\Sigma)$ denote the set
\begin{equation}\label{LepsilonSigma}
\cL_\epsilon(\Sigma) = \pi_\epsilon^{-1}(\Sigma \cap X_\epsilon).
\end{equation}
This enjoys the same properties of the Leray coboundary discussed
above. In particular, notice that $\cL_\epsilon(\partial\Sigma)=\partial
\cL_\epsilon(\Sigma)$.

\smallskip

We consider forms $\pi^*(\eta)$ as in \eqref{etapullback}. To keep
track explicitly of the order of pole of such forms along the
hypersurface $X$, we modify the notation and write
\begin{equation}\label{etam}
\pi^*(\eta_m) = \frac{\Delta(\omega)}{f^m},
\end{equation}
with $\omega$ and $f$ as in \eqref{etapullback}.

\smallskip

We then make the following proposal for a regularization method for
the Feynman integrals \eqref{ParFeySigma2}. We call it {\em Leray
regularization}, because it is based on the use of Leray
coboundaries. (Notice that this procedure of regularization and
subtraction happens after having already removed the divergent
$\Gamma$-factor from the parametric Feyman integrals and passing
to residues. It is meant in fact to take care of the remaining
singularities that arise from the intersections of the hypersurface
with the domain of integration.)

\begin{defn}\label{LerayRegDef}
The Leray regularized Feynman integral is
obtained from \eqref{ParFeySigma2} by replacing the part
\begin{equation}\label{ParFeyepsilon}
\int_{\partial\Sigma\cap D_\epsilon(X)} \pi^*(\eta_m) + 
\int_{\Sigma\cap D_\epsilon(X) }df \wedge \frac{\pi^*(\eta_m)}{f} 
\end{equation}
of \eqref{ParFeySigma2} with the integral
\begin{equation}\label{ParFeyepsilonNew}
\int_{\cL_\epsilon(\partial\Sigma)} \frac{\pi^*(\eta_{m-1})}{f-\epsilon} + 
\int_{\cL_\epsilon(\Sigma)} df \wedge \frac{\pi^*(\eta_m)}{f-\epsilon}. 
\end{equation}
\end{defn}

Thus, the Leray regularization introduced here consists of replacing
the integral over $\Sigma \cap D_\epsilon(X)$ with an integral over 
$\cL_\epsilon (\Sigma)\simeq (\Sigma\cap X_\epsilon) \times S^1$, which
avoids the locus $\Sigma \cap X$ where the divergence can occur by
going around it along a circle of small radius $\epsilon >0$. 

\smallskip

Using the result of Lemma \ref{dfintLem}, we can formulate
\eqref{ParFeyepsilonNew} equivalently in the following form.

\begin{lem}\label{LerayFeyIntLem}
The Leray regularization of the Feynman integral \eqref{ParFeySigma2} can be
equivalently written in the form
\begin{equation}\label{LerayRegUGamma}
\begin{array}{rl}
\bU(\Gamma)_\epsilon = & \displaystyle{\frac{1}{C(n,D,\ell)} 
\left(\int_{\partial\Sigma \cap D_\epsilon(X)^c}
\pi^*(\eta_m) +\int_{\Sigma\cap D_\epsilon(X)^c} df \wedge \frac{\pi^*(\eta_m)}{f} \right)}
\\[5mm] + &  \displaystyle{\frac{2\pi i} {C(n,D,\ell)}  \left(\int_{\partial\Sigma \cap
X_\epsilon} \frac{\pi^*(\eta_{m-1})}{df}  +  \int_{\Sigma\cap X_\epsilon} \pi^*(\eta_m),
\right) } \end{array}
\end{equation}
with $\pi^*(\eta_m)=\Delta(\omega)/f^m$ as in \eqref{etam} and
Proposition \ref{FeynDelta}.
\end{lem}

\proof The result follows directly from Proposition \ref{FeynDelta}
and Lemma \ref{dfintLem} applied to \eqref{ParFeyepsilonNew}.
\endproof

In \eqref{LerayRegUGamma} we use the notation $D_\epsilon(X)^c$ to
denote the complement of $D_\epsilon(X)$. Notice how only the part of
the integral \eqref{ParFeySigma2} that is computed inside $D_\epsilon(X)$
is replaced by \eqref{ParFeyepsilonNew} in the Leray regularization,
while the part of the integral \eqref{ParFeySigma2} computed outside 
of $D_\epsilon(X)$ remains unchanged. 

\smallskip

We now study the dependence on the parameter $\epsilon >0$ of the
Leray regularized Feynman integral \eqref{ParFeyepsilonNew}, that is,
of the integral
\begin{equation}\label{LerayRegUGamma1}
I_\epsilon:= \int_{\partial\Sigma \cap X_\epsilon} \frac{\pi^*(\eta_{m-1})}{df}  + 
\int_{\Sigma\cap X_\epsilon} \pi^*(\eta_m).
\end{equation}

\smallskip

\begin{thm}\label{ancontI}
The function $I_\epsilon$ of \eqref{LerayRegUGamma1} is infinitely
differentiable in $\epsilon$. Moreover, it extends to a holomorphic
function for $\epsilon \in \Delta^*\subset \C$, a small punctured
disk, with a pole of order at most $m$ at $\epsilon =0$, with $m$ as
in \eqref{mnDell}.
\end{thm}

\proof To prove the differentiability of $I_\epsilon$, let us write 
\begin{equation}\label{Aepsilon}
A_\epsilon(\eta) = \int_{\Sigma\cap X_\epsilon} \pi^*(\eta),
\end{equation}
with $\pi^*(\eta)$ as in \eqref{etapullback}. By Lemma \ref{dfintLem} above,
and the fact that $d\pi^*(\eta)=0$, we obtain
\begin{equation}\label{dAepsilon}
\frac{d}{d\epsilon} A_\epsilon(\eta) = - \int_{\partial \Sigma\cap
X_\epsilon} \frac{ \pi^*(\eta)}{df},
\end{equation}
where $f$ is as in \eqref{fnDell} and $\pi^*(\eta)/df$ is the
Gelfand--Leray form of $\pi^*(\eta)$. Thus, we can write
$$ I_\epsilon = A_\epsilon(\eta_m) - \frac{d}{d\epsilon} A_\epsilon(\eta_{m-1}). $$
Thus, to check the differentiability in the variable $\epsilon$ to all
orders of $I_\epsilon$ is equivalent to checking that of
$A_\epsilon$. We define then $\Upsilon:
\Omega^n \to \Omega^n$ by setting
\begin{equation}\label{UpGL}
\Upsilon(\alpha) = d\left(\frac{\alpha}{df}\right),
\end{equation}
where $\alpha/df$ is a Gelfand--Leray form for $\alpha$. 
In turn, the $n$-form $\Upsilon(\alpha)$ also has a Gelfand--Leray form,
which we denote by
\begin{equation}\label{deltaGL}
\delta(\alpha)=\frac{\Upsilon(\alpha)}{df} =
\frac{d\left(\frac{\alpha}{df}\right)}{df}. 
\end{equation}
We then prove that, for $k\geq 2$,
\begin{equation}\label{dkAepsilon}
\frac{d^k}{d\epsilon^k} A_\epsilon =- \int_{\partial\Sigma\cap
X_\epsilon} \delta^{k-1}\left( \frac{\pi^*(\eta)}{df} \right).
\end{equation}
This follows by induction. In fact, we first see that
$$ \frac{d^2}{d\epsilon^2} A_\epsilon = - \frac{d}{d\epsilon} 
\int_{\partial \Sigma\cap X_\epsilon} \frac{ \pi^*(\eta)}{df} $$
which, applying Lemma \ref{dfintLem} gives
$$ = - \int_{\partial \Sigma\cap X_\epsilon} \frac{d\left( \frac{
\pi^*(\eta)}{df}\right)}{df}. $$
Assuming then that 
$$ \frac{d^k}{d\epsilon^k} A_\epsilon =
- \int_{\partial\Sigma\cap
X_\epsilon} \delta^{k-1}\left( \frac{\pi^*(\eta)}{df} \right) $$
we obtain again by a direct application of Lemma \ref{dfintLem} 
$$ \frac{d^{k+1}}{d\epsilon^{k+1}} A_\epsilon =
- \int_{\partial\Sigma\cap
X_\epsilon} \frac{d\left(\frac{\delta^{k-1}\left( \frac{\pi^*(\eta)}{df}\right)}{df}
\right)}{df} $$ $$
= - \int_{\partial\Sigma\cap X_\epsilon} \delta^k\left( \frac{\pi^*(\eta)}{df}\right). $$
This proves differentiability to all orders. 

Notice then that, while the expression \eqref{ParFeyepsilonNew} used
in Definition \ref{LerayRegDef} is, a priori, only defined for
$\epsilon >0$, the equivalent expression given in the second line of
\eqref{LerayRegUGamma} and in \eqref{LerayRegUGamma1} is clearly
defined for any complex $\epsilon \in \Delta^*$ in a punctured disk
around $\epsilon=0$ of sufficiently small radius.
It can then be seen that the expression
\eqref{LerayRegUGamma1} depends holomorphically
on the parameter $\epsilon$ by the general argument on holomorphic
dependence on parameters given in Part III, \S 10.2 of Vol.II of
\cite{AGZV}. 

Finally, to see that $I_\epsilon$ has a pole of order at
most $m$ at $\epsilon=0$, notice that the form $\pi^*(\eta_m)$ of
\eqref{etam} is given by $\Delta(\omega)/f^m$ and has a pole
of order at most $m$ at $X$. This is evident in the two cases with 
$n-\frac{D(\ell+1)}{2} \geq 0$ or $n-\frac{D\ell}{2} \leq 0$. It
also holds in the intermediate case with $n-\frac{D(\ell+1)}{2} <0 <
n-\frac{D\ell}{2}$, since we are taking the convention that, in the
case a hypersurface $X$ defined by a polynomial $f=f_1^{n_1} f_2^{n_2}$, a
form $\Delta(\omega)/f^m$ has pole order $m$ along $X$, even thought
on the individual components it has order $mn_1$ and $mn_2$,
respectively. 
\endproof

\smallskip

In particular, the result of Proposition \ref{ancontI} shows that we
can use the Leray regularization as an alternative to dimensional
regularization to replace a divergent Feynman integral by a
meromorphic function of a complex variable $\epsilon$ with a pole at
$\epsilon=0$. It is then possible to proceed as in dimensional
regularization and apply ``minimal subtraction'', namely subtract the
polar part of the resulting Laurent series in $\epsilon$ and evaluate
the remaining part at $\epsilon=0$.

\smallskip

It is clear that this regularization method is subject to the same
problems as dimensional regularization when it comes to considering 
Feynman integrals associated to graphs that contain subdivergences.
One can organize the hierarchy of subdivergences using the
Bogolyubov-Parashuk preparation, as in the case of dimensional
regularization. 

\medskip

\subsection{Birkhoff factorization and renormalization}\label{BirkhoffSec}

Connes and Kreimer \cite{CoKr} showed that the BPHZ renormalization
procedure, in the DimReg+MS regularization scheme, can be understood
conceptually as the Birkhoff factorization of loops in the Lie group of complex 
points of the affine group scheme $G$ dual to a commutative Hopf algebra $\cH$ 
generated by the Feynman diagrams of the given physical theory. 
The Hopf algebra $\cH$, at the discrete combinatorial level, is the commutative algebra 
generated by the one-particle-irreducible (1PI) graphs of the theory, with the
coproduct
\begin{equation}\label{Hopf}
\Delta(\Gamma) = \Gamma\otimes 1 + 1 \otimes \Gamma + \sum \gamma \otimes \Gamma/\gamma,
\end{equation}
where the sum is over proper subgraphs $\gamma \subset \Gamma$
satisfying a set of properties such as being Feynman diagrams of the
same theory (see for instance 
\cite{CoMa-book} for a detailed discussion of the assumptions on the
family of subgraphs involved in the coproduct). The quotient
$\Gamma/\gamma$ denotes the graph obtained by contracting each
component of $\gamma$ to a single vertex. It is sometimes denoted in
the literature with the notation $\Gamma // \gamma$. The Hopf
algebra is graded by the number of internal lines of graphs.

After identifying loops $\gamma: \Delta^* \to G(\C)$, defined on an infinitesimal
punctured disk $\Delta^*$ around $z=0$, with elements $\phi\in G(K)=\Hom(\cH,K)$,
where $K$ is the field of germs of meromorphic functions at $z=0$, Connes and Kreimer showed
that the BPHZ formula for renormalization is the recursive formula 
\begin{equation}\label{BPHZ}
\begin{array}{l}
\phi_-(x)=- T (\phi(x)+ \sum \phi_-(x') \phi(x'')) \\[2mm]
\phi_+(x)= \phi(x) +\phi_-(x) +\sum \phi_-(x') \phi(x''),
\end{array}
\end{equation}
with $\Delta(x)=x\otimes 1 + 1 \otimes x  +\sum x' \otimes x''$, and
$x',x''$ of lower degree, and with $T$ the projection of a Laurent
series onto its polar part. 
The original BPHZ formula is obtained by applying \eqref{BPHZ} to the
element $\phi \in \Hom(\cH,K)$ 
that assigns to a generator $\Gamma$ of $\cH$ its unrenormalized
Feynman integral $U(\Gamma)$.  
As shown in \cite{CoKr}, the formula \eqref{BPHZ} is the recursive
formula that gives the Birkhoff factorization 
\begin{equation}\label{Birkhoffgamma}
 \gamma(z) =\gamma_-(z)^{-1} \gamma_+(z) 
\end{equation} 
of the loop $\gamma$ into a part $\gamma_+$ that is holomorphic on $\Delta$ and
a part $\gamma_-$ that is holomorphic at $\infty\in \P^1(\C)$, where
one identifies $\gamma_+$ 
with $\phi_+\in \Hom(\cH,\cO)$, with $\cO$ the algebra of germs of
holomorphic functions at $z=0$ 
and $\gamma_-$ with $\phi_- \in \Hom(\cH,\cQ)$ wth $\cQ=\C[z^{-1}]$ so that
\begin{equation}\label{Birkhoffphi}
 \phi = (\phi_- \circ S) * \phi_+, 
\end{equation} 
with $S$ the antipode of $\cH$ and $*$ the product in the affine group
scheme $G$, dual to the coproduct of $\cH$. 

\medskip

The formulation in terms of Birkhoff factorization of loops with
values in the Lie group of complex points of the affine group scheme
of diffeographisms is applied in \cite{CoKr} to the Dimensional
Regularization of Feynman integrals. Namely, the dimensionally
regularized Feynman integrals $U(\Gamma)(z)$ of \eqref{DRint} define
an element $\phi \in \Hom(\cH,K)$, with $\cH$ the Connes--Kreimer Hopf
algebra of Feynman graphs of the theory and $K$ the field of germs of
meromorphic functions at $z=0$, given by assigning as values on the
generators of the Hopf algebra
\begin{equation}\label{phiUGamma}
\phi(\Gamma) = U(\Gamma) \in K.
\end{equation}
In the case of dimensional regularization of Feynman integrals, the 
fact that the $U(\Gamma)(z)$ define meromorphic functions is very
delicate, see the discussion in \S 1.4 of
\cite{CoMa-book}, especially Lemma 1.7, Lemma 1.8, and Theorem 1.9.
On the contrary, we have seen that, using the Leray coboundary regularization
introduced above, one easily obtains meromorphic functions of the
parameter $\epsilon$. We return to discuss the analytic continuation
to meromorphic functions of the dimensionally regularized integrals
via a different approach in \S \ref{MellinSec} below.

\smallskip

By the results of \S \ref{LerRegSec} above, we can apply the same
BPHZ renormalization procedure to the Leray coboundary regularization
introduced in Definition \ref{LerayFeyIntLem}. We thus consider the element
$\phi \in  \Hom(\cH,K)$ defined by assigning on generators 
\begin{equation}\label{phiLeray}
\phi(\Gamma)(\epsilon) =U(\Gamma)_\epsilon
\end{equation}
defined as in \eqref{LerayRegUGamma}. By Proposition \ref{ancontI}, we
know that $U(\Gamma)_\epsilon$ defines a germ of a meromorphic
function for $\epsilon\in \Delta^*$, an infinitesimal punctured disk
around $\epsilon=0$, hence it defines an element in $K$.
We can then apply the Birkhoff factorization of $\phi$, as in
\eqref{Birkhoffphi}. This provides the counterterms, in the form
\begin{equation}\label{LerayCounter}
C(\Gamma)_\epsilon = \phi_-(\Gamma)(\epsilon),
\end{equation}
which, as a function of $\epsilon$, is an element in $\cQ$, and the
renormalized value of the Feynman integral, given by the finite value
at zero
\begin{equation}\label{LerayRenorm}
R(\Gamma) = \phi_+(\Gamma)(0),
\end{equation}
where $\phi_+(\Gamma)(\epsilon)$ defines an element in the ring of
convergent power series $\cO\subset K$.

\medskip

\subsection{Mellin transform and the DimReg integral}\label{MellinSec}

We now return to consider the method of Dimensional Regularization and
reinterpret it in terms of oscillatory integrals and mixed Hodge
structures. 
As we recalled briefly in \S \ref{GLSec} above, the oscillatory
integrals used in the theory of singularities and mixed Hodge
structures can be seen as Fourier transforms \eqref{intGLform}
of a Gelfand--Leray function \eqref{GLfunct}. One can also consider,
instead of a Fourier transform, a Mellin transform of the same 
Gelfand--Leray function. Since Mellin and Fourier transform determine
each other by well known formulae, the information obtained in this
way is equivalent. In the context of singularity theory, the 
Mellin transforms of Gelfand--Leray functions and its relation to the
oscillatory integral is discussed, for instance, in 
Part II, \S 7.2.1, of \cite{AGZV}, Vol.II.

\medskip

It was already proved by Belkale and Brosnan in \cite{BeBro2} that,
in the case of log-divergent graphs, the dimensionally regularized
parametric Feynman integral can be written as a local Igusa L-function.
This was later generalized to the non-log-divergent case in the
work of Bogner and Weinzierl \cite{BoWei1}, \cite{BoWei2}, \cite{BoWei3}.
Our approach here is closely related to these results, though we do not
discuss in detail the explicit relation. Moreover, we simplify the
form of the integrals with respect to the case considered by Bogner
and Weinzierl, so that we do not have to perform the cutting into
sectors and blowups. We rely, in fact, on the formulation in terms
of the exponential of the rational function $V_\Gamma(t,p)$ and its
expansion, and we analyze the resulting terms individually. A more detailed
analysis using the formulation of Bogner--Weinzierl and Belkale--Brosnan
is possible, but we do not consider it here. 

\medskip

In order to relate the dimensionally regularized parametric Feynman
integral to the oscillatory integrals and the Mellin transforms of
Gelfand--Leray functions, consider again the integrals of the form
\eqref{DRint}, or better, the similar integrals computed after slicing
with a $k$-plane $\Pi_\xi$ as in \S \ref{SliceSec}, so that the
intersection $X_\Gamma\cap \Pi_\xi$ has isolated singularities.

As shown in Lemma \ref{DRlem}, we can equivalently compute
the dimensionally regularized Feynman integral \eqref{DRint} 
using the form \eqref{expintDR}. Thus, we first consider an integral
of the form
\begin{equation}\label{intDRexpxi}
\int_{\Pi_\xi^+} \frac{e^{-V_\Gamma(t,p)}}{\Psi_\Gamma(t)^{(D+z)/2}} \,\, \omega_\xi
\end{equation}
where $\Pi_\xi^+=\Pi_\xi\cap \R_+^n$ and $\omega_\xi$ is as in \eqref{omegaxi}.
After expanding the exponential term and using \eqref{ratioVGamma}, we are reduced 
to considering integrals of the form
\begin{equation}\label{intDRellxi}
\int_{\Pi_\xi^+} \frac{P_\Gamma(t,p)^\ell}{\Psi_\Gamma(t)^{\ell+(D+z)/2}}\, \omega_\xi.
\end{equation}

Thus, we concentrate here on integrals of the form
\begin{equation}\label{DRxihf}
F_{\Gamma,\xi}(z) = \int \Psi_\Gamma^z \, \chi_\xi\, P_\Gamma^\ell \Omega_\xi,
\end{equation}
with $\Omega_\xi$ is as in \eqref{Omegaxi}, and for some integer $\ell
\geq 0$. We have made here a simple change of coordinates on the complex
variable $z$, whose meaning will become apparent in a moment.

The function $\chi_\xi$ in \eqref{DRxihf} is the characteristic function
of the domain of integration. In order to show that one can extract from these
dimensionally regularized Feynman integrals information on the singularities
of the graph hypersurface $X_\Gamma$ (through its slices $X_\Gamma
\cap \Pi_\xi$), it suffices to concentrate on the part of the domain
of integration that is close to the hypersurface $X_\Gamma$. Thus, we
can include in the function $\chi_\xi$ an additional cutoff of the
integral that is supported in a neighborhood of the intersection
$\Sigma_\xi \cap X_\Gamma$ of the original domain of integration in $\Pi_\xi$ 
with the graph hypersurface.

In the following, for simplicity of notation, we just write
\eqref{DRxihf} as
\begin{equation}\label{DRxihf2}
F_{\Gamma,\xi}(z) = \int \Psi_\Gamma^z \,  \alpha_\xi,
\end{equation}
where 
\begin{equation}\label{alphaxi}
\alpha_\xi = \chi_\xi \, P_\Gamma^\ell \, \Omega_\xi.
\end{equation}

\medskip

\begin{lem}\label{MellinFGamma}
The function \eqref{DRxihf2} is the Mellin transform of the
Gelfand-Leray function
\begin{equation}\label{GLxiJ}
J_{\Gamma,\xi}(\epsilon)= \int_{X_\epsilon} \frac{\alpha_\xi}{df}, 
\end{equation}
with $f=\Psi_\Gamma|_{\Pi_\xi}$.
\end{lem}

\proof First observe that both functions $\Psi_\Gamma(t)$ and $P_\Gamma(t,p)$
are real when restricted to the domain $\Sigma\subset \R_+^n$, with
$\Psi_\Gamma(t)>0$ on the interior of this domain. Thus, we can write the function 
$F_{\Gamma,\xi}(z)$ of \eqref{DRxihf2} in the form
\begin{equation}\label{Mellin1}
 F_{\Gamma,\xi}(z) =\int_0^\infty s^z \left( \int_{X_s}
\frac{\alpha_\xi}{df} \right) \, ds.  
\end{equation}
One can recognize then that \eqref{Mellin1} is in fact the Mellin
transform 
\begin{equation}\label{Mellin2}
F_{\Gamma,\xi}(z) =\int_0^\infty s^z J_{\Gamma,\xi}(s) \, ds,
\end{equation}
for $J_{\Gamma,\xi}$ as in \eqref{GLxiJ}, the corresponding Gelfand--Leray function.
\endproof

\medskip

The identification of $F_{\Gamma,\xi}(z)$ with the Mellin transform
\eqref{Mellin2} also provides an answer to the problem of the analytic
continuation to meromorphic functions in the complex plane for
functions of the form \eqref{DRxihf}. This analytic continuation is 
needed in order to justify our change of variables in $z$ in passing from
\eqref{intDRellxi} to \eqref{DRxihf}, as well as the use of integrals of the
form \eqref{DRxihf} to derive conclusions about the original
dimensionally regularized integrals \eqref{intDRellxi}.
In fact, the existence of an analytic continuation to meromorphic
functions for the functions $F_{\Gamma,\xi}(z)$ follows from the
existence of an asymptotic expansion for Gelfand--Leray functions of the form
\begin{equation}\label{Jpol}
J(s)= \int_{X_s} \frac{\alpha}{df}, \ \ \  \alpha = h \chi \omega_n,
\end{equation}
with $h$ a polynomial term and $\chi$ a compactly supported smooth
function, supported near an isolated singularity of the hypersurface
$f=0$. The asymptotic expansion is given by
\begin{equation}\label{Jasympt}
J(s) \sim \sum_{\lambda \in \Xi} \sum_{r=0}^{n-1} a_{r,\lambda}\,
s^\lambda \log(s)^r, \ \ \ \ s\to 0^+
\end{equation}
with $\Xi$ a discrete subset of $\R$. The points $\lambda\in \Xi$
depend on the set of multiplicities of an embedded resolution of the
singularity, see Part II, \S 7 of \cite{AGZV} and \cite{Var1}.
This implies the following result (\cf \cite{AGZV}), for generic
choice of the slicing $\Pi_\xi$ and of the external momenta.

\smallskip

\begin{cor}\label{analcontFGamma}
Suppose that the cutoff function $\chi_\xi$ in \eqref{alphaxi} is
supported in a small neighborhood of an isolated singularity of
$X_\Gamma \cap \Pi_\xi$. Then the function $F_{\Gamma,\xi}(z)$, 
defined as in \eqref{DRxihf2} for $\Re(z)>0$ sufficiently large, 
admits an analytic continuation to a
meromorphic function over the whole complex plane, with poles at the 
discrete set of points $z=-(\lambda+1)$, with $\lambda\in\Xi$ as in 
\eqref{Jasympt}, with the coefficient of $(z+\lambda+1)^{-(r+1)}$ in
the Laurent series expansion given by $(-1)^r r! a_{r,\lambda}$, with
$a_{r,\lambda}$ as in \eqref{Jasympt}.
\end{cor}

\medskip

\subsection{Dimensional regularization and mixed Hodge
structures}\label{MHSsec}

We use the results of the previous section relating the
dimensional regularization of the Feynman integrals to the Mellin
transform of Gelfand--Leray functions, and the results of \S
\ref{MilFibSec} on the interpretation in terms of cohomology of the
Milnor fiber, to relate the dimensionally regularized Feynman
integrals to limiting mixed Hodge structures.

\smallskip

We assume here to be in the case of isolated singularities, possibly
after replacing the original Feynman integrals with their slices along 
planes $\Pi_\xi$ of dimension complementary to that of the singular
locus of the hypersurface, as discussed in \S\S \ref{RadonSec} and
\ref{SliceSec} above. 

\smallskip

The cohomological Milnor fibration has fiber over $\epsilon$ 
given by the complex vector space $H^{k-1}(F_\epsilon,\C)$,
where the Milnor fiber $F_\epsilon$ of $X_\xi$ is homotopically a
bouquet of $\mu$ spheres $S^{k-1}$, with $k=\dim \Pi_\xi -1$ and 
with $\mu$ the Milnor number of
the isolated singularity. A holomorphic $k$ form $\alpha = h
\omega_\xi/f^m$ determines a section of the cohomological Milnor
fibration by taking the classes
\begin{equation}\label{secMF}
\left[ \frac{\alpha}{df} |_{F_\epsilon} \right] \in
H^{k-1}(F_\epsilon,\C). 
\end{equation}
We then have the following results (\cite{AGZV}, Vol.II \S 13).
The asymptotic formula \eqref{Jasympt} for the Gelfand--Leray
functions implies that the function of $\epsilon$ obtained by
pairing the section \eqref{secMF} with a locally constant section of
the homological Milnor fibration has an asymptotic expansion
\begin{equation}\label{expMFsec}
\left\langle \left[ \frac{\alpha}{df} \right], \delta \right\rangle \sim
\sum_{\lambda,r} \frac{a_{r,\lambda}}{r!} \epsilon^\lambda \log(\epsilon)^r ,
\end{equation}
for $\epsilon \to 0$, where $\delta(\epsilon)\in
H_{k-1}(F_\epsilon,\Z)$. Moreover, there exist classes
\begin{equation}\label{etaalpha}
\eta^\alpha_{r,\lambda} (\epsilon) \in H^{k-1}(F_\epsilon, \C)
\end{equation}
such that the coefficients $a_{r,\lambda}$ of \eqref{expMFsec} are
given by
\begin{equation}\label{etaalphaint}
\langle \eta^\alpha_{r,\lambda} (\epsilon),\delta(\epsilon) \rangle
=a_{r,\lambda} .
\end{equation}
Thus, one defines the ``geometric section'' associated to the
holomorphic $k$-form $\alpha$ as
\begin{equation}\label{geosec}
\sigma(\alpha):= \sum_{r,\lambda} \eta^\alpha_{r,\lambda}(\epsilon)
\frac{\epsilon^\lambda \log(\epsilon)^r}{r!}.
\end{equation}
The order of the geometric section $\sigma(\alpha)$ is defined as
being the smallest $\lambda$ in the discrete set $\Xi\subset \R$ such
that $\eta^\alpha_{0,\lambda}\neq 0$. One denotes it with
$\lambda_\alpha$. The {\em principal part} of 
$\sigma(\alpha)$ is then defined as
\begin{equation}\label{princsigma}
\sigma_{\max}(\alpha)(\epsilon):= \epsilon^{\lambda_\alpha} \left(
\eta^\alpha_{0,\lambda_\alpha} + \cdots +
\frac{\log(\epsilon)^{k-1}}{(k-1)!} \eta^\alpha_{k-1,\lambda_\alpha} \right),
\end{equation}
where one knows that 
\begin{equation}\label{Neta}
\eta^\alpha_{r,\lambda} = \cN^r \eta^\alpha_{0,\lambda},
\end{equation}
where $\cN$ is the nilpotent operator given by the logarithm of the
unipotent monodromy, given by 
$$ \cN = -\frac{1}{2\pi i} \log \cT  $$
with $\log \cT  =\sum_{r\geq 1} (-1)^{r+1} (\cT - id)^r /r$.

The {\em asymptotic mixed Hodge structure} on the fibers of the
cohomological Milnor fibration constructed by Varchenko (\cite{Var2},
\cite{Var3}) has as the Hodge filtration the subspaces $F^r \subset
H^{k-1}(F_\epsilon,\C)$ defined by
\begin{equation}\label{VarFr}
F^r=\{ [\alpha/df] \,|\, \lambda_\alpha \leq k-r-1 \}
\end{equation}
and as weight filtration $W_\ell \subset H^{k-1}(F_\epsilon,\C)$ the
filtration associated to the nilpotent monodromy operator $\cN$. 
This mixed Hodge structure has the same weight filtration as the
{\em limiting mixed Hodge structure} constructed by Steenbrink
(\cite{Steen}, \cite{Steen2}), but the Hodge filtration is different,
though the two agree on the graded pieces of the weight filtration. 

\medskip

We now use a refined version of the results of \S \ref{MilFibSec},
and in particular Corollary \ref{MilFibH} for Feynman integrands 
as in \eqref{xiformsDell}. We show that, upon varying the choice of
the external momenta $p$ and of the spacetime dimension $D$, the
corresponding Feynman integrands, in a neighborhood of an isolated
singular point of $X_\Gamma \cap \Pi_\xi$, determine a subspace of the
cohomology $H^{k-1}(F_\xi,\C)$ of the Milnor fiber of $X_\Gamma \cap
\Pi_\xi$. This inherits a Hodge and a weight filtrations from the
Milnor fiber cohomology with its asymptotic mixed Hodge structure.
We concentrate on the case where $k-D\ell/2 \leq 0$, so that we can
consider, for fixed $k$, arbitrarily large values of $D\in \N$.

\begin{prop}\label{subHMilnor}
Consider Feynman integrals, sliced along a linear space $\Pi_\xi$ as
in \eqref{UGammaxi}. We write the integrand in the form 
\begin{equation}\label{alphaxiagain}
 \alpha_\xi = \frac{h \Omega_\xi}{f^m}, 
\end{equation}
with 
\begin{equation}\label{hfmkD}
\left\{  \begin{array}{l} 
h = P_\Gamma^{-k+D\ell/2} \\[2mm]
f = \Psi_\Gamma \\[2mm]
m=-k+D(\ell+1)/2,
\end{array}\right.
\end{equation}
as in \eqref{alpham3}, with $k-D\ell/2 \leq 0$. 
Upon varying the external momenta $p$ in $P_\Gamma(p,t)$ and the
spacetime dimension $D\in \N$, with $k-D\ell/2 \leq 0$, the forms
$\alpha_\xi$ as above determine a subspace 
$$ H_{{\rm Feynman}}^{k-1}(F_\epsilon,\C) \subset
H^{k-1}(F_\epsilon,\C), $$
of the fibers of the cohomological Milnor fibration, spanned by
elements of the form \eqref{alphaxiagain},
where the polynomials $h=h_{T,v,w,p}$ are of the form   
\begin{equation}\label{talphaMilH}
h(t) = \prod_{i=1}^{-k+D\ell/2} L_{T_i}(t) \prod_{e\notin T_i} t_e,
\end{equation}
where the $T_i$ are spanning trees and the $L_{T_i}(t)$ are the linear
functions of \eqref{hTpolyn}.
\end{prop}

\proof Consider the explicit
expression \eqref{PGammapt} of the polynomial $P_\Gamma(t,p)$ as a
function of the external momenta, through the coefficients $s_C$ of 
\eqref{sCcoeff}. 
One can see that, by varying arbitrarily the external momenta,
subject to the global conservation law \eqref{conslawpe}, one can
reduce to the simplest possible case, where all external momenta are
zero except for a pair of opposite momenta $P_{v_1}=p=-P_{v_2}$
associated to a pair of external edges attached 
to a pair of vertices $v_1,v_2$. In such a case, the polynomial 
$P_\Gamma(t,p)$ becomes of the form \eqref{vwPGamma}. Thus, when
considering powers $P_\Gamma(t,p)^{-k+D\ell/2}$ for varying $D$, 
we obtain all polynomials of the form \eqref{talphaMilH}.
\endproof

\smallskip

We denote by $H^{k-1}_{{\rm Feynman}}(F_\epsilon,\C)$ the subspace of
the cohomology $H^{k-1}(F_\epsilon,\C)$ of the Milnor fiber spanned by
the classes $[\alpha_\xi/df]$ with $\alpha_\xi$ of the form
\eqref{alphaxiagain}, with $h$ of the form \eqref{talphaMilH},
considered modulo the ideal generated by the derivatives of
$f=\Psi_\Gamma$ and localized at an isolated singular point, \ie
viewed as elements in the Milnor algebra $M(f)$.
The subspace $H^{k-1}_{{\rm Feynman}}(F_\epsilon,\C)$ inherits a Hodge
and a weight filtration $F^\bullet \cap H^{k-1}_{{\rm Feynman}}$ and
$W_\bullet \cap H^{k-1}_{{\rm Feynman}}$ from the asymptotic mixed Hodge
structure of Varchenko on $H^{k-1}(F_\epsilon,\C)$. It is an
interesting problem to see whether
the subspace $H^{k-1}_{{\rm Feynman}}$ recovers
the full $H^{k-1}(F_\epsilon,\C)$ and if $(F^\bullet \cap H^{k-1}_{{\rm
Feynman}}, W_\bullet \cap H^{k-1}_{{\rm Feynman}})$ still give a mixed
Hodge structure, at least for some classes of graphs $\Gamma$.

\medskip

\section{Regular and irregular singular connections}\label{RegIrregSec}

An important and still mysterious aspect of the motivic approach to
Feynman integrals and renormalization is the problem of reconciling
the Riemann--Hilbert correspondence of perturbative renormalization
formulated by Connes--Marcolli in \cite{CoMa} (see also
\cite{CoMa-book}), which is based on equivalence classes of certain
{\em irregular singular} connections, with the setting of motives
(especially mixed Tate motives) and mixed Hodge structures, which are
naturally related to {\em regular singular} connections.  The
irregular singular connections of \cite{CoMa} have values in the Lie
algebra of the Connes--Kreimer group of diffeographisms and are
defined on a fibration over a punctured disk with fiber the
multiplicative group, respectively representing the complex variable
$z$ of dimensional regularization and the energy scale $\mu$ (or
rather $\mu^z$) upon which the dimensionally regularized Feynman
integrals depend. On the other hand, in the case of hypersurfaces in
projective spaces, the natural associated regular singular connection
is the Gauss--Manin connection on the cohomology of the Milnor fiber
and the Picard--Fuchs equation for the vanishing cycles. We sketch
here a relation between this regular singular connection and the
irregular equisingular connections of \cite{CoMa}. (To avoid any
possible confusion, the reader should
keep in mind that the use of the term ``equisingular'' in \cite{CoMa}
is not the same as the well established use in singularity theory, as
in \cite{Teiss} for instance.)

\subsection{Picard--Fuchs equation and Gauss--Manin connection}

In the following we let
\begin{equation}\label{basisvanishing}
\left[\frac{\omega_i}{df}\right] \ \ \ \  i=1,\ldots,\mu
\end{equation}
be a basis for the vanishing cohomology bundle, written with the same
notation we used above for the Gelfand--Leray form. Then the
Gauss--Manin connection on the vanishing cohomology bundle, which is
defined by the integer cohomology lattice in each real cohomology
fiber, acts on the basis \eqref{basisvanishing} by 
\begin{equation}\label{actionGM}
\nabla^{GM}_s \left[\frac{\omega_i}{df}\right]_s = \sum_j p_{ij}(s)
\left[\frac{\omega_j}{df}\right]_s, 
\end{equation} 
where the $p_{ij}(s)$ are holomorphic away from $s=0$ and have a pole
at $s=0$. The Gauss--Manin connection is regular singular and its
monodromy agrees with the monodromy of the singularity (see
\cite{AGLV}, \S 2.3). Given a covariantly constant section $\delta(s)$
of the vanishing homology bundle, 
the function 
\begin{equation}\label{It}
 I(s)=\left( \int_{\delta(s)} \frac{\omega_1}{df}, \ldots,
\int_{\delta(s)} \frac{\omega_\mu}{df} \right) 
\end{equation}
is a solution of the regular-singular Picard--Fuchs equation 
\begin{equation}\label{PFeq}
\frac{d}{ds} I(s) = P(s) I(s), \ \ \ \ \text{ with } \ \ P(s)_{ij}=p_{ij}(s).
\end{equation}

\medskip

Similarly, suppose given a holomorphic $n$-form $\omega$ and let
$\omega/df$ be the corresponding Gelfand--Leray form, defining a
section $[\omega/df]$ of the vanishing cohomology bundle. Let
$\delta_1, \ldots, \delta_\mu$ be a basis of the vanishing homology,
$\delta_i(s) \in H_{n-1}(F_s,\Z)$. Then the function
\begin{equation}\label{Is2}
I(s)=\left( \int_{\delta_1(s)} \frac{\omega}{df}, \ldots,
\int_{\delta_\mu(s)} \frac{\omega}{df} \right) 
\end{equation}
satisfies a regular singular order $\ell$ differential equation
\begin{equation}\label{PellDE}
I^{(\ell)}(s)+p_1(s) I^{(\ell-1)}(s)+\cdots+p_\ell(s) I(s)=0,
\end{equation}
where the order is bounded above by the multiplicity of the critical
point (see \cite{AGZV}, \S 12.2.1). One refers to \eqref{PellDE}, or
to the equivalent system of regular singular homogeneous first order equations 
\begin{equation}\label{sysPell}
\frac{d}{ds} \cI(s)= \cP(s) \cI(s),
\end{equation}
with 
\begin{equation}\label{cIs}
\cI_r(s)=s^{r-1} I^{(r-1)}(s),
\end{equation}
as the Picard--Fuchs
equation of $\omega$. For the relation between Picard--Fuchs equations
and mixed Hodge structures see \S 12 of \cite{AGZV} and
\cite{Kulikov}. 

\medskip

\subsection{Flat equisingular connections}\label{EqGrIndSec}

We first recall some properties of the flat
equisingular connections introduced in \cite{CoMa} (see also
 \S 1 of \cite{CoMa-book}). We denote by $G$ the affine group scheme
dual to the commutative Hopf algebra of Feynman diagrams, graded by
loop number. We let $\cg$ denote the Lie algebra $\cg=\Lie(G)$. Let
$K$ denote the field of germs of meromorphic functions at $z=0$. We
also let $B$ denote a fibration over an infinitesimal disk $\Delta^*$
with fiber the multiplicative group $\bG_m$ and we denote by $P$ the
principal $G$-bundle $P=B\times G$. We consider $\Lie(G)$-valued flat
connections $\omega$ that are {\em equisingular}, \ie they satisfy 
\begin{itemize}
\item The connections satisfies $\omega(z,\lambda u)=\lambda^Y \omega(z,u)$, for
$\lambda\in \bG_m$, with $Y$ the grading operator. 
\item Solutions of $D\gamma=\omega$,  
have the property that their pullbacks
$\sigma^*(\gamma)\in G(K)$ along any section $\sigma: \Delta \to B$ with fixed
value $\sigma(0)$ have the same negative piece of the Birkhoff
factorization $\sigma^*(\gamma)_-$.
\end{itemize}

The first condition and the flatness condition imply that the
connection $\omega(z,u)$ can be written in the form
\begin{equation}\label{omegazu}
\omega(z,u) = u^Y(a(z))\, dz + u^Y(b(z))\, \frac{du}{u} ,
\end{equation}
where $a(z)$ and $b(z)$ are elements of $\cg(K)$ satisfying the
flatness condition
\begin{equation}\label{flat}
\frac{db}{dz} - Y(a) +[a,b] =0.
\end{equation}

Recall that the Lie bracket in the Lie algebra $\Lie(G)$ is obtained by assigning
\begin{equation}\label{Liegraphs}
[ \Gamma, \Gamma' ] = \sum_{v\in V(\Gamma)} \Gamma \circ_v \Gamma' -
\sum_{v'\in V(\Gamma')} \Gamma'\circ_{v'}\Gamma,
\end{equation}
where $\Gamma \circ_v \Gamma'$ denotes the graph obtained by inserting
$\Gamma'$ into $\Gamma$ at the vertex $v\in V(\Gamma)$ and the sum is
over all vertices where an insertion is possible.

The equisingularity condition, which determines the behavior of
pullbacks of solutions along sections of the fibration $\bG_m \to B
\to \Delta$, can be checked by writing the equation
$Df=\omega$ in the more explicit form
\begin{equation}\label{Eqfomega}
\gamma^{-1} \frac{d\gamma}{dz} = a(z), \ \ \ \text{ and } \ \ \   
\gamma^{-1} Y(\gamma)= b(z). 
\end{equation}
When one interprets elements $\gamma \in G(K)$ as algebra
homomorphisms $\phi \in \Hom(\cH,K)$, one can write the above
equivalently in the form
\begin{equation}\label{Eqfomega1}
(\phi \circ S) * \frac{d\phi}{dz} = a, \ \ \ \text{ and } \ \ \   
(\phi\circ S) *  Y(\phi)= b, 
\end{equation}
where $S$ is the antipode in $\cH$ and $*$ is the product dual to the
coproduct in the Hopf algebra. This means, on generators $\Gamma$ of $\cH$,
\begin{equation}\label{Eqfomega2}
\langle (\phi \circ S) \otimes \frac{d\phi}{dz}, \Delta(\Gamma)
\rangle = a_\Gamma, \ \ \ \text{ and } \ \ \   
\langle(\phi\circ S) \otimes  Y(\phi), \Delta(\Gamma)\rangle= b_\Gamma, 
\end{equation}
where
$$ \Delta(\Gamma)=\Gamma\otimes 1 + 1\otimes \Gamma + \sum_\gamma \gamma \otimes \Gamma/\gamma $$
as in \eqref{Hopf}, with the sum over subdivergences, and the antipode
is given inductively by
\begin{equation}\label{antipode}
S(X)= -X -\sum S(X') X'', 
\end{equation}
for $\Delta(X)=X\otimes 1 + 1 \otimes X + \sum X'\otimes X''$, with
$X'$ and $X''$ of lower degree.

\medskip

\subsection{From regular to irregular
singularities}\label{RegtoIrrSec} 

We now show how to produce a flat connection of the desired form
\eqref{omegazu}, with irregular singularities, starting from the 
graph hypersuraces $X_\Gamma$, a consistent choice of slicing 
$\Pi_\xi$, and the regular singular Picard--Fuchs equation 
associated to the resulting isolated singularities of $X_\Gamma \cap
\Pi_\xi$. 

\smallskip

We begin by introducing a small modification of the Hopf algebra and
coproduct, which accounts for the fact of having to choose a slicing
$\Pi_\xi$. This is similar to what happens when one enriches the
discrete Hopf algebra by adding the data of the external momenta.

\smallskip

Let $\cS_\Gamma$ denote the manifold of planes $\Pi_\xi$ in
$\A^{\#E(\Gamma)}$ with $\dim \Pi_\xi \leq \codim\, {\rm
Sing}(X_\Gamma)$. We can write $\cS_\Gamma$ as a disjoint union
\begin{equation}\label{mSGamma}
\cS_\Gamma = \displaystyle{\bigcup_{m=1}^{\codim\, {\rm Sing}(X_\Gamma)} \cS_{\Gamma,m}},
\end{equation}
where $\cS_{\Gamma,m}$ is the manifold of $m$-dimensional planes 
in $\A^{\#E(\Gamma)}$. We denote by $\cC^\infty(\cS_\Gamma)$ the space
of test functions on $\cS_\Gamma$ and by $\cC^{-\infty}_c(\cS_\Gamma)$
its dual space of distributions.

\smallskip

\begin{lem}\label{sigmainduce}
Suppose given a subgraph $\gamma\subset \Gamma$. Then the choice of 
a distribution $\sigma \in \cC^{-\infty}_c(\cS_\Gamma)$ induces
distributions $\sigma_\gamma \in \cC^{-\infty}_c(\cS_\gamma)$ and
$\sigma_{\Gamma/\gamma} \in \cC^{-\infty}_c(\cS_{\Gamma/\gamma})$.
\end{lem}

\proof Given $\gamma \subset \Gamma$, neglecting external edges, we
can realize the affine $X_\gamma$ as a hypersurface inside a linear 
subspace $\A^{\# E(\gamma)}\subset \A^{\# E(\Gamma)}$ and similarly
for the affine $X_{\Gamma/\gamma}$, seen as a hypersurface inside 
a linear subspace $\A^{\# E(\Gamma/\gamma)}\subset \A^{\# E(\Gamma)}$,
where we simply identify the edges of $\gamma$ or $\Gamma/\gamma$
with a subset of the edges of the original graph $\Gamma$. 

One then has a restriction map $T_\gamma : \cS_{\Gamma,\gamma} \to
\cS_\gamma$, where $\cS_{\Gamma,\gamma} \subset \cS_\Gamma$ is the
union of the components $\cS_{\Gamma,m}$ of $\cS_\Gamma$ with
$m\leq \codim\, {\rm Sing}(X_\gamma)$,
\begin{equation}\label{SgammaGamma}
\cS_{\Gamma,\gamma} = \bigcup_{m=1}^{\codim\, {\rm Sing}(X_\gamma)}\cS_{\Gamma,m},
\end{equation}
which is given by
\begin{equation}\label{Tgamma}
T_\gamma(\Pi_\xi)= \Pi_\xi\cap \A^{\# E(\gamma)}.
\end{equation}
This induces a map $T_\gamma: \cC^\infty(\cS_\gamma) \to
\cC^\infty(\cS_\Gamma)$ given by
\begin{equation}\label{Tgammaf}
T_\gamma(f)(\Pi_\xi)=\left\{ \begin{array}{ll} f(T_\gamma(\Pi_\xi)) & \Pi_\xi
\in \cS_{\Gamma,\gamma} \\[2mm]
0 & \text{otherwise.}
\end{array}\right.
\end{equation}
In turn, this defines a map $T_\gamma: \cC^{-\infty}_c(\cS_\Gamma) \to
\cC^{-\infty}_c(\cS_\gamma)$, at the level of distributions, by
\begin{equation}\label{Tgammasigma}
T_\gamma(\sigma)(f)=\sigma(T_\gamma(f)).
\end{equation}
The argument for $\Gamma/\gamma$ is analogous. One sets $\sigma_\gamma
= T_\gamma(\sigma_\Gamma)$ and $\sigma_{\Gamma/\gamma}
= T_{\Gamma/\gamma}(\sigma_\Gamma)$.
\endproof

We then enrich the original Hopf algebra $\cH$ by adding the datum of
the slicing $\Pi_\xi$. We consider the commutative algebra
\begin{equation}\label{tildeH}
\tilde\cH ={\rm Sym}(\cC^{-\infty}_c (\cS)),
\end{equation}
where $\cS=\cup_\Gamma \cS_\Gamma$, endowed with the coproduct
\begin{equation}\label{tildecoprod}
\Delta(\Gamma,\sigma)= (\Gamma,\sigma)\otimes 1 + 1\otimes
(\Gamma,\sigma) +\sum_\gamma (\gamma,\sigma_\gamma)\otimes
(\Gamma/\gamma,\sigma_{\Gamma/\gamma}). 
\end{equation}

\begin{lem}\label{coass}
The coproduct \eqref{tildecoprod} is coassociative and $\tilde\cH$ is a Hopf algebra.
\end{lem}

\proof The proof is analogous to the one given in \cite{CoMa-book},
Theorem 1.27.
\endproof

\medskip

We then proceed as follows. We pass to the projective instead of
affine formulation and we fix a small neighborhood of an isolated 
singular point of $X_\Gamma \cap \Pi_\xi$, for $\Pi_\xi$ a linear 
space of dimension at most equal to the codimension of ${\rm
Sing}(X_\Gamma)$. Suppose given a holomorphic $k$-form $\alpha_\xi$ on
$\Pi_\xi$. Then there exists an associated regular singular 
Picard--Fuchs equation  
\begin{equation}\label{PFJxi}
J_{\Gamma,\xi}^{(\ell)}(s) + p_1(s)J_{\Gamma,\xi}^{(\ell-1)}(s)+
\cdots + p_\ell(s) J_{\Gamma,\xi}(s)   =0,
\end{equation}
with the property that any solution $J_{\Gamma,\xi}(s)$ is a linear
combination of the functions
\begin{equation}\label{intalphaxi}
J_{\Gamma,\xi,i}(s)=\int_{\delta_i(s)} \frac{\alpha_\xi}{df},
\end{equation}
where $\delta_1,\ldots,\delta_\mu$ be a basis of locally constant sections
of the homological Milnor fibration, $\delta_i(s) \in
H_{k-1}(F_s,\Z)$, and $\alpha_\xi/df$ is the Gelfand--Leray form
associated to the holomorphic $k$-form $\alpha_\xi$.

\smallskip

This depends on the choice of a singular point and can be localized in
a small neighborhood of the singular point in $X_\Gamma \cap
\Pi_\xi$. In fact, introducing a cutoff $\chi_\xi$ as in
\eqref{DRxihf} that is supported near the singularities of 
$X_\Gamma \cap \Pi_\xi$ amounts to adding the expressions
\eqref{intalphaxi} for the different singular points. Thus, to
simplify notations, we can just assume to have a single expression
$J_{\Gamma,\xi}(s)$ at a unique isolated critical point. 

\smallskip

We then have the following result, which constructs irregular singular
connections as in \S \ref{EqGrIndSec} from solutions of the regular
singular  Picard--Fuchs equation.

\begin{thm}\label{JGammaEquising}
Any solution $J_{\Gamma,\xi}$ of the regular singular Picard--Fuchs
equation \eqref{PFJxi} determines a flat $\cg(K)$-valued connection
$\omega(z,u)$ of the form \eqref{omegazu}. Moreover, if the $k$-form
$\alpha_\xi$ is given by $P_\Gamma^\ell \Omega_\xi$ as in
\eqref{alphaxi}, then the connection is equisingular. 
\end{thm}

\proof  We consider the Mellin transform, as in \eqref{Mellin2}
\begin{equation}\label{iMellinJxi}
\cF_{\Gamma,\xi}(z)= \int_0^\infty s^z \,J_{\Gamma,\xi}(s)\, ds.
\end{equation}
As in Corollary \ref{analcontFGamma} (see \S 7 of \cite{AGZV}), the
function $\cF_{\Gamma,\xi}(z)$ admits an analytic continuation to
meromorphic functions with poles at points $z=-(\lambda+1)$ with
$\lambda\in \Xi_{\Gamma,\xi}$ a discrete set in $\R$ of points related
to the multiplicities of an embedded resolution of the singular point
of $X_\Gamma\cap \Pi_\xi$. We look at the function
$\cF_{\Gamma,\xi}(z)$ in a small neighborhood of a chosen point
$z=-D$. It has an expansion as a Laurent series, with a pole at 
$z=-D$ if $-D \in \Xi_{\Gamma,\xi}$. 

After a change of variables on the complex coordinate $z$, so that we have
$z\in\Delta^*$ a small neighborhood of $z=0$, we define
\begin{equation}\label{phiFGamma}
\phi_\mu(\Gamma,\sigma)(z):= \mu^{-z \,b_1(\Gamma)} \sigma\left(
\cF_{\Gamma,\xi}( -\frac{D+z}{2}) \right),   
\end{equation}
where we consider $\cF_{\Gamma,\xi}$ as a function of $\xi$ to which
we apply the distribution $\sigma$. More precisely, after identifying 
$\cF_{\Gamma,\xi}$ with its Laurent series expansion, we apply $\sigma$ to
the coefficients seen as functions of $\xi$. This defines an algebra
homomorphism $\phi_\mu\in \Hom(\tilde\cH,K)$, by assigning the values
\eqref{phiFGamma} on generators. Here $\mu$ is the mass scale as in \S
\ref{MassSec} above. The homomorphism $\phi$ defined by
\eqref{phiFGamma} can be equivalently described as a family of 
$\tilde G(\C)$-valued loops $\gamma_\mu: \Delta^* \to \tilde G(\C)$, depending on
the mass scale $\mu$. Here $\tilde G$ denotes the affine group scheme
dual to the commutative Hopf algebra $\tilde\cH$.
The dependence on $\mu$ of \eqref{phiFGamma}
implies that $\gamma_\mu$ satisfies the scaling property
\begin{equation}\label{scalegammamu}
\gamma_{e^t \mu}(z) =\theta_{tz}(\gamma_\mu(z)), 
\end{equation}
where $\theta_t$ is the one-parameter family of automorphisms of
$\tilde\cH$ generated by the grading, $\frac{d}{dt}\theta_t|_{t=0}=Y$.
Then one sets
\begin{equation}\label{abphi}
a_\mu(z):= (\phi_\mu \circ S)* \frac{d}{dz}\phi_\mu , \ \ \ \text{ and } \
\ \ b_\mu(z):= (\phi_\mu \circ S)* Y(\phi_\mu),
\end{equation}
where $S$ and $*$ are the antipode of $\tilde\cH$ and the product dual
to the coproduct $\Delta$ of \eqref{tildecoprod}. These define
elements $a_\mu$, $b_\mu$ $\Omega^1(\cg(K))$, which one can use 
to define a connection $\omega(z,u)$ of the form \eqref{omegazu}.
More precisely, for $\mu=e^t$, one has
$$ \gamma_\mu^{-1}\frac{d}{dz}\gamma_\mu =\theta_t(\gamma^{-1}
\frac{d}{dz}\gamma)= u^Y(a(z)), $$
where we set $u^Y=e^{tY}$ and then extend the resulting expression to
$u\in \bG_m(\C)=\C^*$. Similarly, we get $\gamma_\mu^{-1}
Y(\gamma_\mu)= u^Y(b(z))$. Thus, the connection $\omega(z,u)$ defined
in this way satisfies by construction the first condition of the
equisingularity property, namely $\omega(z,\lambda u)=\lambda^Y
\omega(z,u)$, for all $\lambda \in \bG_m$. One can see that the
connection is flat since we have
$$ \frac{d}{dz}b_\mu(z)- Y(a_\mu(z))=  \frac{d\gamma_\mu^{-1}(z)}{dz}
Y(\gamma_\mu(z)) + \gamma_\mu^{-1}(z) \frac{d}{dz}(Y(\gamma_\mu(z)))
$$ $$ -Y(\gamma_\mu^{-1}(z))\frac{d}{dz}\gamma_\mu(z) -
\gamma_\mu^{-1}(z) \frac{d}{dz}(Y(\gamma_\mu(z)))  $$ 
$$ = -\gamma_\mu^{-1}(z) \frac{d}{dz}(\gamma_\mu(z))
\gamma_\mu^{-1}(z) Y(\gamma_\mu(z)) - \gamma_\mu^{-1}(z)
Y(\gamma_\mu(z)) \gamma_\mu^{-1}(z) =- [a(z),b(z)]. $$

The second condition of
equisingularity is the property that, in the Birkhoff
factorization 
$$ \gamma_\mu(z) = \gamma_{\mu,-}(z)^{-1} \gamma_{\mu,+}(z), $$
the negative part satisfies
$$ \frac{d}{d\mu} \gamma_{\mu,-}(z) =0. $$
By dimensional analysis on the counterterms, in the case of
Dimensional Regularization and Minimal Subtraction, it is possible to
show (see \cite{Collins} \S 5.8.1) that the counterterms obtained by
the BPHZ procedure applied to the Feynman integral
$U_\mu(\Gamma)(z)$ of \eqref{expintDRmu} and \eqref{muDRint}
do not depend on the mass parameter $\mu$. This
means, as shown in \cite{CoKr} (see also Proposition 1.44 of
\cite{CoMa-book}), that the Feynman integrals $U_\mu(\Gamma)(z)$
define a $G(\C)$-valued 
loop $\gamma_\mu(z)$ with the property that $\partial_\mu
\gamma_{\mu,-}(z) =0$. The integrals \eqref{iMellinJxi} considered
here, in the case where $\alpha_\xi$ is of the form \eqref{alphaxi},
correspond to slices along a linear space $\Pi_\xi$ of the Feynman
integrals \eqref{expintDRmu}, localized by a cutoff $\chi_\xi$ near
the singular points. The explicit dependence on $\mu$ in
the integrals \eqref{ParFeySigmaxi} is as in \eqref{phiFGamma}, which 
is unchanged with respect to that of the original dimensionally 
regularized Feynman integrals \eqref{expintDRmu}. Thus, the same
argument of \cite{Collins} \S 5.8.1 and Proposition 1.44 of
\cite{CoMa-book} applies to this case to show that $\partial_\mu
\gamma_{\mu,-}(z) =0$.
\endproof

\medskip

\section{Logarithmic motives, Dimensional Regularization, and motivic sheaves}

In this section we propose a candidate for a motivic formulation of
dimensional regularization. As we discussed already in \S
\ref{DimRegSec} above, in physics dimensional
regularization is intended as a purely formal recipe that assigns a
meaning to Gaussian integrals in ``complexified dimension" $z\in \C$
by continuation to non-integer values of the usual formula for integer
dimensions 
\begin{equation}\label{IntGaussDimReg}
\int e^{-\lambda t^2} d^z t := \pi^{z/2} \lambda^{-z/2}.
\end{equation}

Usually, in so doing, one does not attempt to give a geometric meaning
to the space of integration as a ``space in complexified dimension
$z\in \C$''. The question of whether one can actually make sense of a
geometry in complexified dimension was considered in \cite{CoMa-book},
from the point of view of noncommutative geometry, where the usual
notion of dimension of a space is replaced by the dimension spectrum,
which is a set of complex numbers. A geometric model for a space whose
dimension spectrum consists of a single point $z\in \C^*$ is given in 
\S I.19.2 of \cite{CoMa-book}, and it is shown that the formula
\eqref{IntGaussDimReg} can be recovered from the properties of the
Dirac operator on this space.

Here we also consider the question of giving geometric meaning to the
complexified dimension, but we try to construct a geometric model
underlying the operation of dimensional regularization using motives.
We propose a candidate for a motive describing the dimensional
regularization of a given Feynman graph. This is defined as an
extension (in fact as a pro-motive) in the category of mixed motives,
which is obtained from the logarithmic extension of Tate motives and
the motive of the graph hypersurface. 
We work in the geometric setting of motivic sheaves. 
One can choose to work in a similar way at the level of Hodge structures,
using Hodge sheaves. For a formulation of some aspects of dimensional
regularization in a setting that is closer to that of mixed Hodge modules,
we refer the reader to \S 7 of \cite{Mar-book} and the unpublished \cite{CoMa-anom}. 

Just as in the case of
noncommutative geometry, where the operation of dimensional
regularization is understood as a product of the ordinary space in
integer dimension by the ``space of dimension $z$'', here we also find
that the dimensionally regularized Feynman integral is recovered by
taking the product, in a category of motivic sheaves, of the motive
associated to the graph hypersurface of a given Feynman graph by this
pro-motive representing the ``space of dimension $z$''.
It would be interesting to find a more explicit relation between this 
motivic description of dimensional regularization and the one based on
noncommutative geometry, described in \cite{CoMa-book} and \cite{CoMa-anom}.

\subsection{Mixed Tate motives and the logarithmic extensions}

We recall briefly the definition of the logarithmic motives, as given
in \cite{Ayoub}.  Let $\cD\cM(\bG_m)$ be the Voevodsky category of
mixed motives (motivic sheaves) over the multiplicative group
$\bG_m$. We will assume that the base field $\K$ is a number field (in
fact, we can work over $\Q$) so that the extensions considered here
take place in an abelian category of mixed Tate motives (\cf
\cite{Andre}, \cite{Levine}).  
Recall that the extensions $\Ext^1_{\cD\cM(\K)}(\Q(0),\Q(1))$ of Tate
motives are given by the Kummer motives $M=[\Z\stackrel{u}{\to}
\bG_m]$ with $u(1)=q\in \K^*$. This extension has period matrix of the form 
\begin{equation}\label{periodKummer} 
\left( \begin{array}{cc} 1 &  0 \\
\log q & 2\pi i \end{array} \right).
\end{equation}

When, instead of working with motives over the base field $\K$, one
works with the relative setting of motivic sheaves over a base scheme
$S$, instead of the Tate motives $\Q(n)$ one considers the Tate
sheaves $\Q_S(n)$. These correspond to the constant sheaf with the
motive $\Q(n)$ over each point $s\in S$. In the case where
$S=\bG_m$, there is a natural way to assemble the Kummer motives
into a unique extension in
$\Ext^1_{\cD\cM(\bG_m)}(\Q_{\bG_m}(0),\Q_{\bG_m}(1))$. This is the 
Kummer extension 
\begin{equation}\label{Kummer}
\Q_{\bG_m}(1) \to \cK \to \Q_{\bG_m}(0) \to \Q_{\bG_m}(1)[1],
\end{equation}
where over the point $s\in \bG_m$  one is taking the Kummer
extension $M_s=[\Z \stackrel{u}{\to} \bG_m]$ with $u(1)=s$.  
Because of the logarithm function $\log (s)$ that appears in the
period matrix for this extension, the Kummer extension
\eqref{Kummer} is also referred to as the {\em logarithmic motive}. We
use the notation $\cK = {\rm Log}$ as in \cite{Ayoub} to refer to this
extension, \cf \cite{BeDe}.

When working with $\Q$-coefficients, so that one can include
denominators in the definition of projectors, one can then consider
the logarithmic motives ${\rm Log}^n$, defined as in \cite{Ayoub} by
setting 
\begin{equation}\label{Logn}
{\rm Log}^n = {\rm Sym}^n(\cK),
\end{equation}
where the symmetric powers of an object in $\cD\cM_\Q(\bG_m)$ are defined as
\begin{equation}\label{SymnX}
{\rm Sym}^n(X) = \frac{1}{\# \Sigma_n} \sum_{\sigma\in \Sigma_n} \sigma(X^n).
\end{equation}

Recall that the polylogarithms appear naturally as period matrices for extensions involving the
symmetric powers ${\rm Log}^n={\rm Sym}^n(\cK)$, in the form \cite{BloMM}
\begin{equation}\label{PolylogMot}
0 \to {\rm Log}^{n-1}(1) \to \cL^n \to \Q(0) \to 0,
\end{equation}
where $M(1)=M\otimes \Q(1)$ and $\cL^1={\rm Log}$. The mixed motive $\cL^n$ has period matrix
\begin{equation}\label{PolylogMHS}
\left(\begin{array}{cc} 1 & 0 \\ M_{\rm Li}^{(n)} & M_{{\rm Log}^{n-1}(1)} \end{array}\right) 
\end{equation}
with 
\begin{equation}\label{PolylogMHS2}
M_{\rm Li}^{(n)} = ( -{\rm Li}_1(s), -{\rm Li}_2(s), \cdots, -{\rm Li}_n(s) )^\tau, 
\end{equation}
where $\tau$ means transpose and where
$$ {\rm Li}_1(s)=-\log(1-s), \ \ \ \text{ and }  \ \ \ {\rm
Li}_n(s)=\int_0^s {\rm Li}_{n-1}(u) \, \frac{du}{u},  $$ 
equivalently defined (on the principal branch) using the power series 
$$ {\rm Li}_n(s) = \sum_k \frac{s^k}{k^n}, $$
and with
\begin{equation}\label{PolylogMHS3}
M_{{\rm Log}^n(1)} = \left( \begin{array}{ccccc} 2\pi i & 0 & 0 & \cdots & 0 \\
2\pi i \log(s) & (2\pi i)^2 & 0 & \cdots & 0 \\
2\pi i \frac{\log^2(s)}{2!} & (2\pi i)^2 \log(s) & (2\pi i)^3 & \cdots & 0 \\
\vdots & \vdots & \vdots & \cdots & \vdots \\
2\pi i  \frac{\log^n(s)}{n!} & (2\pi i)^2 \frac{\log^{n-1}(s)}{(n-1)!}
& (2\pi i)^3 \frac{\log^{n-2}(s)}{(n-2)!} & 
\cdots & (2\pi i)^n 
\end{array}\right).
\end{equation}

The period matrices for the motives ${\rm Log}^n$ correspond to the
description of ${\rm Log}^n$ as extension of $\Q(0)$ by ${\rm
Log}^{n-1}(1)$, \ie to the distinguished triangles in $\cD\cM(\bG_m)$

of the form
\begin{equation}\label{LognExt}
{\rm Log}^{n-1}(1) \to {\rm Log}^n \to \Q(0) \to {\rm Log}^{n-1}(1)[1] .
\end{equation}

\smallskip

The motives ${\rm Log}^n$ form a projective system under the canonical
maps $$\beta_n: {\rm Log}^{n+1}\to {\rm Log}^n$$ given by the
composition of the morphisms ${\rm Sym}^{n+m}(\cK)\to {\rm
Sym}^n(\cK)\otimes {\rm Sym}^m(\cK)$, as in \cite{Ayoub}, Lemma 4.35,
given by the fact that ${\rm Sym}^{n+m}(\cK)$ is canonically a direct
factor of ${\rm Sym}^n(\cK)\otimes {\rm Sym}^m(\cK)$, and the map
${\rm Sym}^m(\cK)\to \Q(0)$ of \eqref{LognExt}, in the particular case
$m=1$. 
Let ${\rm Log}^\infty$ denote the pro-motive obtained as the
projective limit 
\begin{equation}\label{Loginfty}
{\rm Log}^\infty = \varprojlim_n {\rm Log}^n . 
\end{equation}

The analog of the period matrix \eqref{PolylogMHS3} becomes then the infinite matrix
\begin{equation}\label{MHSinfMat}
M_{{\rm Log}^\infty(1)} = \left( \begin{array}{cccccc} 2\pi i & 0 & 0 & \cdots & 0 & \cdots \\
2\pi i \log(s) & (2\pi i)^2 & 0 & \cdots & 0 & \cdots \\
2\pi i \frac{\log^2(s)}{2!} & (2\pi i)^2 \log(s) & (2\pi i)^3 & \cdots & 0 & \cdots \\
\vdots & \vdots & \vdots & \cdots & \vdots & \cdots \\
2\pi i  \frac{\log^n(s)}{n!} & (2\pi i)^2 \frac{\log^{n-1}(s)}{(n-1)!}
& (2\pi i)^3 \frac{\log^{n-2}(s)}{(n-2)!} & 
\cdots & (2\pi i)^n & \cdots \\
\vdots & \vdots & \vdots & \cdots & \vdots & \cdots
\end{array}\right).
\end{equation}

\medskip

In other words, the mixed Hodge structure associated to the motives
${\rm Log}^n$ is the one that has as the weight filtrations $W_{-2k}$
the range of multiplication by the matrix $M_{{\rm Log}^n}$ defined as
in \eqref{PolylogMHS3} on vectors in $\Q^n$ with the first $k-1$
entries equal to zero, while the Hodge filtration $F^{-k}$ is given by
the range of multiplication of $M_{{\rm Log}^n}$ on vectors of $\C^n$
with the entries from $k+1$ to $n$ equal to zero \cite{BloMM}. 

Thus, in this Hodge realization, the $H^0$ piece corresponds to the first
column of the matrix $M_{{\rm Log}^n}$, where the $k$-th entry corresponds
to the $k$-th graded piece of the weight filtration. Let us consider
the corresponding grading operator, that multiplies the $k$-th entry
by $T^k$. One can then associate to the $h^0$-piece of the ${\rm
Log}^\infty$ motive the following formal expression that corresponds
in the period matrix \eqref{MHSinfMat} to the $H^0$ part in the MHS
realization:
\begin{equation}\label{h0Log}
\Q\cdot \sum_k \frac{\log^k(s)}{k!} T^k =: \Q\cdot s^T.
\end{equation}

The formal expression \eqref{h0Log} has in fact an interpretation in
terms of periods. This follows from a well known result (\cf \eg
\cite{GonMTM}, Lemma 2.10) expressing the powers of the logarithm in
terms of iterated integrals.
For iterated integrals we use the notation as in \cite{GonMTM}
\begin{equation}\label{iterint}
\int_a^b \frac{ds}{s} \circ \frac{ds}{s} \circ \cdots \frac{ds}{s}=
\int_{a\leq s_1\leq \cdots \leq s_n \leq b} \frac{ds_1}{s_1}\wedge
\cdots \wedge \frac{ds_n}{s_n}. 
\end{equation}
We also denote by $\Lambda_{a,b}(n)$ the domain
\begin{equation}\label{Lambdanab}
\Lambda_{a,b}(n)=\{ (s_1,\ldots,s_n)\,|\, a\leq s_1\leq \cdots \leq
s_n \leq b \}.
\end{equation}

\begin{lem}\label{logint}
The expression \eqref{h0Log} is obtained as rational multiples of the pairing
\begin{equation}\label{tTperiod}
s^T = \int_{\Lambda_{1,s}(\infty)} \eta(T),
\end{equation}
with $\Lambda_{1,s}(\infty)=\cup_n \Lambda_{1,s}(n)$ and the form 
\begin{equation}\label{etaT}
\eta(T):=\sum_n \frac{ds_1}{s_1}\wedge
\cdots \wedge \frac{ds_n}{s_n}\,\, T^n.
\end{equation}
\end{lem}

\proof The result follows from the basic identity (\cf 
\cite{GonMTM}, Lemma 2.10)
\begin{equation}\label{intlotba}
\int_{\Lambda_{a,b}(n)} \frac{ds_1}{s_1}\wedge
\cdots \wedge \frac{ds_n}{s_n} = \frac{\log\left(\frac{b}{a}\right)^n}{n!}.
\end{equation}
\endproof

\subsection{Motivic sheaves and graph hypersurfaces}

Arapura constructed in \cite{Arapura} a category of motivic sheaves
over a base scheme $S$, modeled on Nori's approach to the construction
of categories of mixed motives. We discuss briefly how a similar
formalism may be applied to the Feynman motives associated to the
graph hypersurfaces with the corresponding periods of the form
\eqref{UGammaParam}. 

\smallskip

The category of motivic sheaves constructed in \cite{Arapura} is based
on Nori's construction of categories of motives via representations of
graphs made of objects and morphisms (\cf \cite{Bru}). In Arapura's
case, one constructs a category of motivic sheaves over a scheme $S$,
by taking as vertices of the corresponding graph objects of the form
\begin{equation}\label{AraObj}
\left( f: X \to S, Y, i, w \right),
\end{equation}
where $f: X \to S$ is a quasi-projective morphism, $Y\subset X$ is a
closed subvariety, $i\in \N$, and $w\in \Z$. One thinks of such an
object as determining a motivic version $h^i_S(X,Y)(w)$ of the local
system given by the (Tate twisted) fiberwise cohomology of the pair
$H^i_S(X,Y;\Q)=R^i f_* j_! \Q_{X\smallsetminus Y}$, where
$j=j_{X\smallsetminus Y} : X\smallsetminus Y\hookrightarrow X$ is the
open inclusion, \ie the sheaf defined by
$$ U\mapsto H^i(f^{-1}(U),f^{-1}(U)\cap Y;\Q). $$
The edges are given by the geometric morphisms, \ie morphisms of
varieties over $S$,
\begin{equation}\label{geomorph}
(f_1: X_1 \to S, Y_1,i,w) \to (f': X_2\to S, Y_2=F(Y),i,w), \ \ \ \text{ with } \
\ \  f_2\circ F = f_1; 
\end{equation}
the connecting morphisms
\begin{equation}\label{connmorph}
(f: X \to S, Y,i+1,w) \to (f|_Y: Y\to S,Z,i,w), \ \ \ \text{ for } \
\ \  Z\subset Y\subset X; 
\end{equation}
and the twisted projection morphisms
\begin{equation}\label{projmorph}
(f: X\times \P^1 \to S, Y\times \P^1 \cup X\times \{0\},i+2,w+1) \to (f: X\to S,Y,i,w). 
\end{equation}

\smallskip

The product in the category of motivic sheaves of \cite{Arapura} is
given by the fibered product
\begin{equation}\label{prodfiber}
\begin{array}{ll}
(X\to S,Y,i,w)\times (X'\to S,Y',i',w')= \\[2mm] (X\times_S X' \to S,
Y\times_S X' \cup X\times_S Y', i+i',w+w').
\end{array}
\end{equation}
This has the following effect on period computations.

\smallskip

\begin{lem}\label{periodProd}
Suppose then given $\Sigma \subset X$ and $\Sigma'\subset X'$,
defining relative homology cycles for $(X,Y)$ and $(X',Y')$,
respectively. One then has, for the fibered product \eqref{prodfiber}, the period pairing
\begin{equation}\label{intfiberprod2}
\int_{\Sigma\times_S \Sigma'} \pi^*_X(\omega) \wedge \pi^*_{X'}(\eta)
= \int_\Sigma  \omega\wedge f^*
f'_*(\eta), 
\end{equation}
where $f:\Sigma\to S$ and $f':\Sigma'\to S$ are the restrictions of
the maps $X\to S$ and $X'\to S$.
\end{lem}

\proof First recall that, when integrating a differential
form on a fibered product, one has the formula
\begin{equation}\label{intfiberprod}
\int_{X\times_S X'} \pi^*_X(\omega) \wedge \pi^*_{X'}(\eta)=\int_{X}
\omega\wedge (\pi_X)_* \pi^*_{X'}(\eta) = \int_X  \omega\wedge f^*
f'_*(\eta), 
\end{equation}
which corresponds to the diagram 
\begin{eqnarray}
\diagram 
& X\times_S X' \dlto^{\pi_X} \drto_{\pi_{X'}} & \\
X \drto^f & & X' \dlto_{f'} \\
& S & 
\enddiagram
\label{fibprod}
\end{eqnarray}
Suppose then given $\Sigma \subset X$ such that $\partial\Sigma
\subset Y$ and $\Sigma'\subset X'$ with $\partial \Sigma'\subset Y'$. 
One has $$ \partial(\Sigma\times_S \Sigma')=\partial\Sigma\times_S
\Sigma' \cup \Sigma \times_S \partial \Sigma' \subset Y\times_S X'
\cup X\times_S Y', $$
so that $\Sigma\times_S \Sigma'$ defines a relative homology class in 
$(X\times_S X', Y\times_S X'\cup X\times_S Y')$. Given elements
$[\omega]\in H^\cdot_S(X,Y)$ and $[\eta]\in H^\cdot_S(X',Y')$, we then apply the
formula \eqref{intfiberprod} to the integration on $\Sigma\times_S
\Sigma'$ and obtain \eqref{intfiberprod2}.
\endproof

\medskip

\subsection{Logarithmic Feynman motives}

Consider then the graph polynomial $\Psi_\Gamma(s)=\det(M_\Gamma(s))$. By
removing the set of zeros of $\Psi_\Gamma$, \ie the graph hypersurface
$X_\Gamma$, we can consider $\Psi_\Gamma$ as a morphism 
\begin{equation}\label{PhiGammaS}
\Psi_\Gamma : \A^{\#E_\Gamma} \smallsetminus \hat X_\Gamma \to \bG_m.
\end{equation}

We can then consider the pullback of the logarithmic motive ${\rm
Log}\in \cD\cM(\bG_m)$ by this morphism, as in the construction of the
logarithmic specialization system given in \cite{Ayoub}.  
This gives a motive
\begin{equation}\label{LogGamma}
{\rm Log}_\Gamma := \Psi_\Gamma^* ({\rm Log}) \in \cD\cM(U_\Gamma),
\end{equation}
where $U_\Gamma=\A^{\#E_\Gamma} \smallsetminus \hat X_\Gamma$.

\smallskip

In fact, a more sophisticated approach would involve considering 
here the ``log complex'' as in \S 9.2 of \cite{Levine2}, \cf also \S
9.4 of \cite{Levine2}, see also \cite{Gon}.

\smallskip

In the context of the category of motivic sheaves of Arapura recalled
above, we can define the Feynman motives as follows. 

\begin{defn}\label{FeynSheaves}
The category of Feynman motivic sheaves, for a fixed scalar quantum
field theory, is the subcategory of the Arapura
category of motivic sheaves over $\bG_m$ spanned by the objects of the
form 
\begin{equation}\label{graphObj}
(\Psi_\Gamma : \A^{\# E(\Gamma)}\smallsetminus \hat X_\Gamma \to \bG_m, \Lambda\smallsetminus
(\Lambda\cap \hat X_\Gamma),\# E(\Gamma) -1 , \# E(\Gamma) -1),
\end{equation}
where $\Gamma$ ranges over the Feynman graphs of the given scalar field
theory, and where
\begin{equation}\label{DeltaChain}
\Lambda = \{ t\in \A^{\# E(\Gamma)}\,|\, \prod_i t_i =0 \}
\end{equation}
is the union of the coordinate hyperplanes.
\end{defn}

The above correspond to the local systems
\begin{equation}\label{locsysGamma}
H^{n-1}_{\bG_m} (\A^n\smallsetminus \hat X_\Gamma, \Lambda\smallsetminus
(\Lambda\cap \hat X_\Gamma), \Q(n-1)),
\end{equation}
with $n=\# E_{int}(\Gamma)$. 

\smallskip

One can also include  as part of the data the slicing by all possible
$k$-dimensional linear spaces $\Pi_\xi\subset \A^{\# E(\Gamma)}$,
with $k\leq \codim\,{\rm Sing} (X_\Gamma)$, as we did in our previous
discussions, and consider instead of the \eqref{graphObj} objects of the form 
\begin{equation}\label{graphObjPi}
(\Psi_\Gamma|_{\Pi_\xi} :\Pi_\xi \smallsetminus (\hat X_\Gamma\cap \Pi_\xi)
\to \bG_m, (\Lambda\cap \Pi_\xi)\smallsetminus 
(\Lambda\cap \hat X_\Gamma\cap \Pi_\xi), k -1 , w).
\end{equation}

\smallskip

\begin{rem}\label{algtopsimplex}{\rm
The reason for taking the cohomology \eqref{locsysGamma} relative to the {\em algebraic
simplex} $\Lambda$, that is, the union of the coordinate hyperplanes
defined by \eqref{DeltaChain} is that, in this way, we can
regard the {\em topological simplex} $\Sigma=\{ t\in
\R_+^n\,|\, \sum_{i=1}^n t_i =1 \}$ as defining a homology
cycle, since $\partial \Sigma\subset \Lambda$.}\end{rem}

\medskip

\subsection{Dimensional Regularization and motives}

In these terms, the procedure of dimensional regularization can then
be described as follows. Consider again the logarithmic (pro)motive,
viewed itself as a motivic sheaf $X_{{\rm Log}^\infty}\to \bG_m$ 
over $\bG_m$. One can then take the product of a Feynman motive
$$ (\Psi_\Gamma:\A^n\smallsetminus \hat X_\Gamma\to \bG_\m , \Lambda \smallsetminus
(\Lambda\cap \hat X_\Gamma), k-1, k-1), $$ 
or more generally one of the form \eqref{graphObjPi}, by the
(pro)motive 
\begin{equation}\label{Logmotsheaf}
 (X_{\rm Log}^\infty\to \bG_m,\Lambda_\infty,0,0),
\end{equation}
where $\Lambda_\infty$ is such that the domain of integration
$\Lambda_{1,t}(\infty)$ of the period computation of Lemma
\ref{logint} defines a cycle. 
The product is given by a fibered product as in \eqref{prodfiber}, namely
\begin{eqnarray}
\diagram \Psi_\Gamma^*({\rm Log}^\infty)= (\A^n \smallsetminus
\hat X_\Gamma)\rtimes_{\bG_m} X_{{\rm Log}^\infty} \rto\dto & X_{{\rm Log}^\infty} \dto
\\ \A^n \smallsetminus \hat X_\Gamma \rto^{\Psi_\Gamma} & \bG_m 
\enddiagram
\label{fiberprod}
\end{eqnarray}

We then have the following interpretation of the dimensionally
regularized Feynman integrals. 

\begin{prop}\label{DimRegMot}
The dimensionally regularized Feynman integral $F_{\Gamma,\xi}(z)$ of
\eqref{DRxihf2} are periods on the product, in the category of motivic
sheaves enlarges to include projective limits, of the Feynman motive
\eqref{graphObjPi} by the logarithmic pro-motive ${\rm Log}^\infty$
seen as the motivic sheaf \eqref{Logmotsheaf}.
\end{prop}

\proof
Consider the product \eqref{fiberprod}, with the two projections
$$ \begin{array}{ll}
\pi_X: (\Pi_\xi \smallsetminus (\hat X_\Gamma\cap
\Pi_\xi))\times_{\bG_m} X_{{\rm Log}^\infty} \to \Pi_\xi \smallsetminus (\hat X_\Gamma\cap
\Pi_\xi)  \\[2mm]
\pi_L: (\Pi_\xi \smallsetminus (\hat X_\Gamma\cap
\Pi_\xi))\times_{\bG_m} X_{{\rm Log}^\infty} \to X_{{\rm Log}^\infty}.
\end{array} $$
and the form $\pi_X^*(\alpha_\xi)\wedge \pi_L^*(\eta(T))$, where
$\alpha_\xi$ is as in \eqref{alphaxi}, and $\eta(T)$ is the 
form on $X_{{\rm Log}^\infty}$ that gives the period \eqref{tTperiod}.
The period computation of Lemma \ref{logint} gives
\begin{equation}\label{perintPsiGamma}
\Psi_\Gamma^* \left( \int_{\Lambda_{1,s}(\infty)} \eta(T)\right) =
\int_{\Lambda_{1,\Psi_\Gamma (t)}(\infty) } \eta(T) = \sum_n
\frac{\log(\Psi_\Gamma(t))^n}{n!} T^n = \Psi_\Gamma(t)^T.
\end{equation}
We then have, by \eqref{intfiberprod2}, 
$$ \int_{(\Sigma \cap \Pi_\xi) \times_{\bG_m} \Lambda_{1,\Psi_\Gamma(t)}(\infty)}
\pi_X^*(\alpha_\xi)\wedge \pi_L^*(\eta(T)) = \int_{\Sigma \cap \Pi_\xi}
\alpha_\xi\wedge (\pi_X)_*\pi_L^*(\eta(T)) = \int_{\Sigma \cap \Pi_\xi} 
\Psi_\Gamma^T \alpha_\xi. $$
This is the integral \eqref{DRxihf2}, up to replacing the formal
variable $T$ of \eqref{h0Log} with the complex DimReg variable $z$.
\endproof

The interpretation that emerges from this calculation is that
performing the dimensional regularization of a Feynman integral can be
thought of as taking the product in the category of motivic sheaves of
the motive (motivic sheaf) of the graph hypersurface by the projective
limit of the logarithmic motives. The variable $z\in \bG_m$ that gives the
complexified dimension of dimensional regularization corresponds to
the 1-parameter group generated by the grading operator associated to
the weight filtration of the logarithmic motives. The dimensionally
regularized integral is then a period of this product motive.

\medskip

\subsection{Motivic zeta function and the DimReg integral}

Kapranov introduced a notion of motivic zeta function by defining
\begin{equation}\label{MotZeta}
Z_X(T) := \sum_{n\geq 0} {\rm Sym^n}(X) \, T^n,
\end{equation}
where the ${\rm Sym}^n(X)$ can be regarded as objects in an abelian
category of motives, or as classes $[{\rm Sym}^n(X)]$ in the
corresponding Grothendieck ring. Kapranov proved that, when $X$ is the
motive of a curve, then the zeta function is a rational function, in
the sense that, given a motivic measure $\mu: K_0(\cM) \to A$, the
zeta function $Z_{X,\mu}(T)\in A[[T]]$ is a rational function of $T$. 
Later, Larsen and Lunts showed that in general this is not true in the
case of algebraic surfaces \cite{LaLu}.  
 
Here we consider the motivic zeta function of the pullback of the
logarithmic motive along the function 
$\Psi_\Gamma$ as in \eqref{PhiGammaS}. Namely, we consider the motivic
zeta function 
\begin{equation}\label{ZetaLogGamma}
Z_{{\rm Log},\Gamma}(T):= \sum_{n\geq 0} {\rm Sym^n}({\rm Log}_\Gamma) \, T^n.
\end{equation}

An interesting question, which we do not address in the present paper, is whether
one can define a motivic lift of the Dimensional Regularization of the Feynman
integral associated to a Feynman graph $\Gamma$ using the motivic zeta
function \eqref{ZetaLogGamma}. In other words, whether one can obtain the zeta function
\begin{equation}\label{ZetaGammaLogMHS}
Z_\Gamma(T):= \sum_{n\geq 0} \frac{\log^n \Psi_\Gamma}{n!} T^n = \Psi_\Gamma^T
\end{equation}
and the associated integrals
\begin{equation}\label{ZetaInts}
\sum_{n\geq 0} \left(\int_{\Sigma \cap \Pi_\xi}  \frac{\log^n \Psi_\Gamma}{n!} \alpha_\xi\right) T^n=
\int_{\Sigma \cap \Pi_\xi}  \Psi_\Gamma^T \alpha_\xi
\end{equation}
in a natural way from the motivic zeta function \eqref{ZetaLogGamma} of $\Psi_\Gamma^*({\rm Log})$. 
We hope to return to this and related questions in following work.

\end{document}